\documentclass[%
 reprint,
 amsmath,amssymb,
 aps,
prb,
]{revtex4-2}
\usepackage{dcolumn}
\usepackage{bm}
\usepackage{siunitx} 
\usepackage{hyperref}
\usepackage{graphicx}
\usepackage{epstopdf}
\usepackage{rotating}
\usepackage{array}
\usepackage{rotating}
\usepackage{slashed}
\usepackage{multirow}

\def\v#1{{\bf#1}}

\def\be{\begin{equation}}
\def\ee{\end{equation}}
\def\bea{\begin{eqnarray}}
\def\eea{\end{eqnarray}}

\def\bfig{\begin{figure}}
\def\efig{\end{figure}}
\def\bcen{\begin{center}}
\def\ecen{\end{center}}
\def\bfl{\begin{flalign}}
\def\efl{\end{flalign}}

\newcommand{\bfsigma}{\mbox{\boldmath$\sigma$\unboldmath}}

\newcommand{\bfPhi}{\mbox{\boldmath$\Phi$\unboldmath}}

\def\ie{{\it i.e.\,}}

\def\lcal{\mbox{$\cal L\,$}}

\def\rcal{\mbox{$\cal R$}}
\def\zcal{\mbox{$\cal Z$}}
\def\<{\langle}
\def\>{\rangle}

\begin{document}


\title{Tight-Binding realization of non-abelian gauge fields: singular spectra and wave confinement}


\author{Y. Hern\'andez-Espinosa}
\affiliation{Universidad Nacional Aut\'onoma de M\'exico, Instituto de F\'isica, Apartado Postal, 04510 Ciudad de M\'exico}

\author{E. Sadurn\'i}
\affiliation{Benem\'erita Universidad Aut\'onoma de Puebla, Instituto de F\'isica, Apartado Postal J-48, 72570 Puebla, M\'exico}


\date{\today}

\begin{abstract}
We present a geometric construction of a lattice that emulates the action of a gauge field on a fermion. The construction consists of a square lattice made of polymeric sites, where all clustered atoms are identical and represented by potential wells or resonators supporting one bound state. The emulation covers both abelian and non-abelian gauge fields. In the former case, Hofstadter's butterfly is reproduced by means of a chain made of rotating dimers, subject to periodic boundary conditions parallel to the chain. A rigorous map between this model and Harper's Hamiltonian is derived. In the non-abelian case, band mixing and wave confinement are obtained by interband coupling using SU(2) as an internal group, \ie the effects are due to non-commutability of field components. A colored model with SU(3) made of trimers is also studied, finding thereby the appearance of flat bands in special configurations. This work constitutes the first all-geometric emulation of the Peierls substitution, and is valid for many types of waves.
\end{abstract}


\maketitle

\tableofcontents

\begin{figure}[t]
\begin{tabular}{c}
\includegraphics[width=0.45\textwidth]{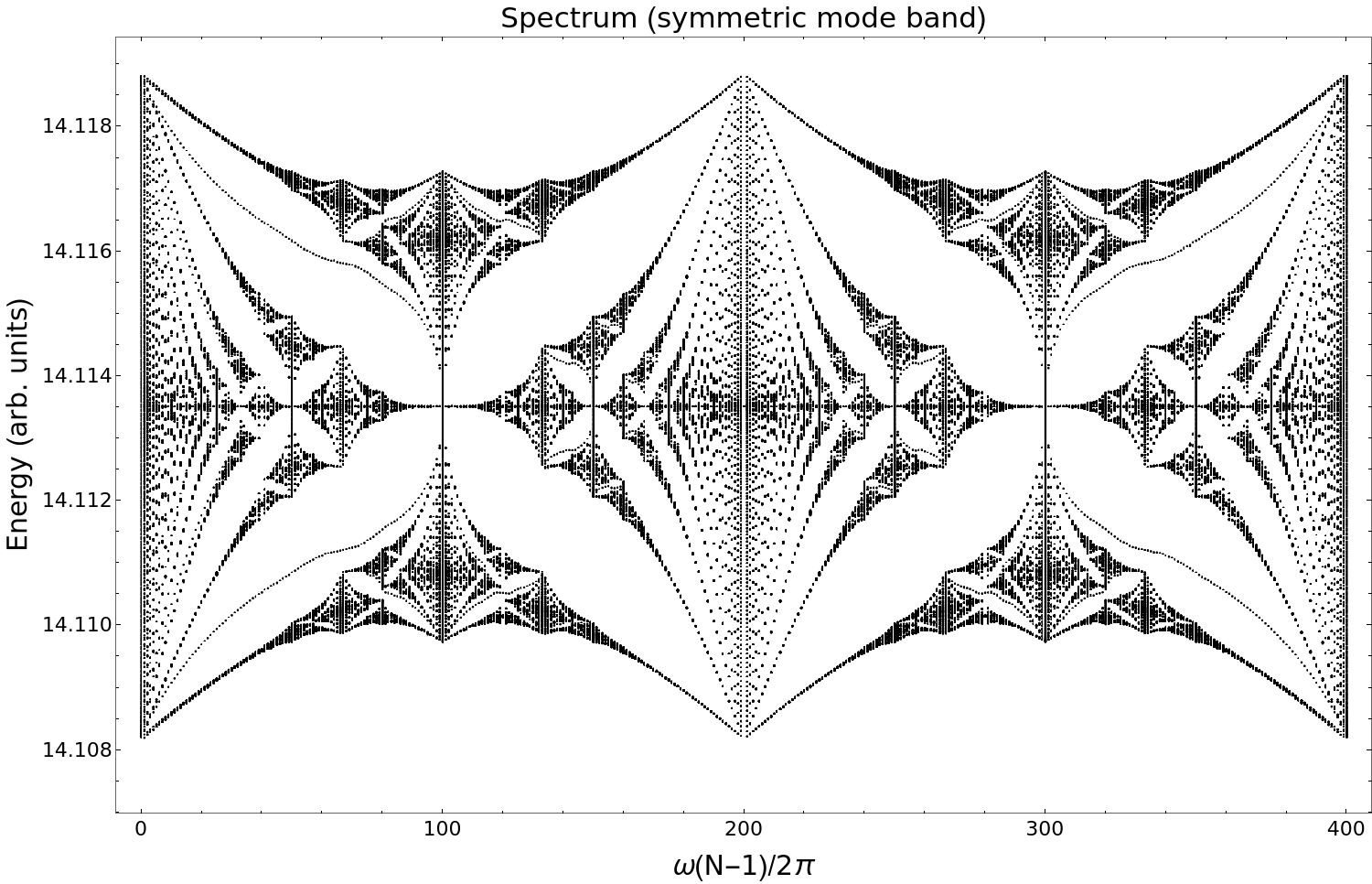}\\
\\
{\large (a)}\\
\\
\includegraphics[width=0.45\textwidth]{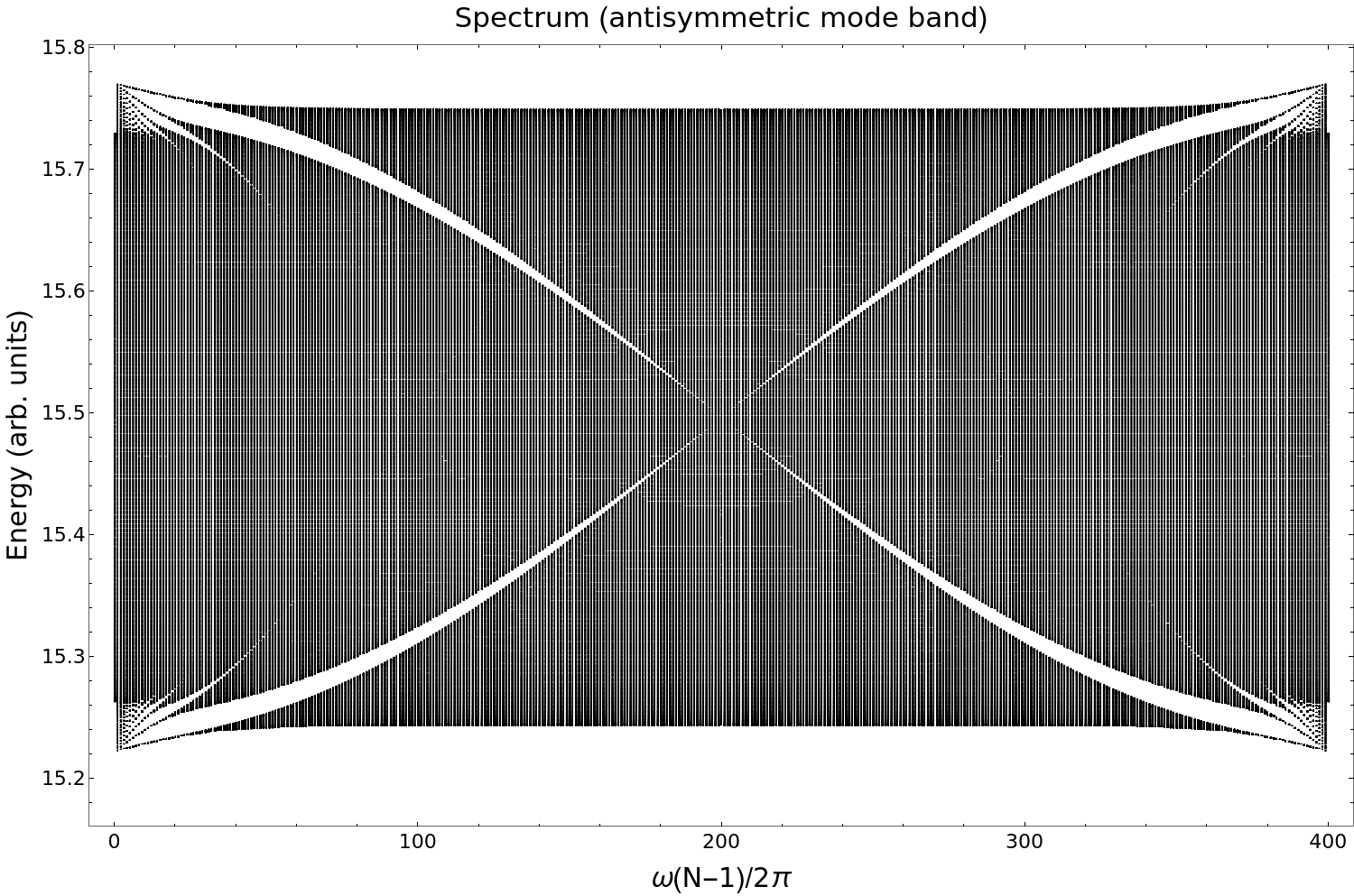}\\
\\
{\large (b)}
\end{tabular}
\caption{\label{fig1} (a) The spectrum of the decoupled symmetric band for a chain of 801 dimers. The Hofstadter butterfly emerges in the first quarter of the frequency  range $\omega(N-1)/2\pi$. The geometric values are $L=1.66$, $d=0.1$ and $\lambda = 1.66$ (b). The spectrum of the decoupled antisymmetric band where the Harper's Hamiltonian is emulated for $\Lambda= 0.08\cos\omega$.The geometric values are $L=2.8$, $d=0.35$ and $\lambda = 1$. Energy is in arbitrary units.}
\end{figure}

\section{\label{introduction} Introduction}

The realization of Yang-Mills theories in lattices is the ultimate goal of spectral quantum emulations \cite{tagliacozzo2013}. The standard model of particles and fields rests on the assumption that internal degrees of freedom such as color are dynamically responsible for the emergence of composite structures in the form of baryons and mesons \cite{greiner}. Lattice QCD is indeed a dynamical description of such theories, where the numerical work is meant to show the emergence of color confinement\cite{epple2005} and infrared slavery \cite{taichiro1979,frasca2014,frasca2009}. However, the implementation of gauge symmetry on the lattice must be done with a proper generalization of Peierls' substitution that incorporates internal degrees of freedom such as color or isospin. In this paper we take care of this aspect, as the lattice emulations presented are sufficiently simple to illuminate important aspects of gauge fields that can be implemented by purely geometrical manipulations of sites on the crystal.

Previous steps towards this goal are the use of atomic levels in experiments involving atomic fountains and the interaction of atoms with the radiation field \cite{kasevich_rf_1989,clairon_ramsey_1991}. We also find optical realizations of abelian gauge fields --e.g. the typical electromagnetic field-- where photons in coupled waveguides \cite{bienstman_taper_2003} play the role of a Schr\"odinger wave under the action of an external force. Complex hopping amplitudes have also been obtained by means of laser techniques \cite{PhysRevLett.108.225303}, and internal degrees of freedom in coupled waveguides have been used with similar purposes \cite{PhysRevA.87.012309}. However, as far as we can see, there has been no attempt to produce these models on a homogeneous lattice without any other assumption than a single quantum state per site. We also take this opportunity to look at microwave realizations with dielectric resonators supporting a single trapped mode. These have been experimentally successful in the emulation of solid state systems such as graphene and boron nitride, which constitute yet another mesoscopic testbed for relativistic spectral emulations \cite{dreisow_classical_2010,hernandez-espinosa_stabilizer_2016, gomes_designer_2012, struck_tunable_2012, fallani_ultracold_2007, bellec_tight-binding_2013, bittner_observation_2010, franco-villafane_first_2013} using the celebrated analogy between the Dirac equation and structures with Dirac points \cite{katsnelson2006,Geim_2007,Castro_Neto_2009,katsnelson2007,katsnelson_zitterbewegung_2006}. 

Our task now is to emulate gauge fields using real scalar waves, starting with the abelian case. We shall see, on the fly, that the procedure giving rise to complex couplings can be employed to produce non-abelian theories as well. With this, we take a step further in our quest for universal emulations using a single-level atom and purely geometrical manipulations. 

We focus on the fermionic part of a non-abelian Yang-Mills theory, where the bosonic part is treated as a static field. The Yang-Mills Lagrangian for vector fields interacting with fermions is:
\bea\label{yangmills}
\lcal=\bar{\psi}i\gamma^\mu\partial_\mu\psi-m\bar{\psi}\psi+ \bar{\psi}\gamma^\mu g A_\mu^a\tau_a\psi-\frac{1}{4}(F^a_{\mu\nu})^2\\
F_{\mu\nu}=\partial_\mu A_\nu-\partial_\nu A_\mu-ig[A_\mu,A_\nu], \quad F_{\mu\nu}=F_{\mu\nu}^a\tau_a.
\eea
As is well-known, the Dirac lagrangian $\bar{\psi}i\gamma^\mu\partial_\mu\psi-m\bar{\psi}\psi$ has been emulated in 2+1 dimensions using hexagonal lattices of one species (graphene, massless case) and two species (boron nitride, massive case). For simplicity we pursue an emulation of non-abelian fields acting on a fermion hopping on a square lattice. Therefore, our attention is focused on the first and second terms of (\ref{yangmills}), but this time formulated as a Peierls non-abelian gauge field. The internal degrees of freedom due to the gauge group will be introduced by means of polymeric sites, i.e. clusters of single-level atoms regarded as a single cell in a crystal. In this respect, the term $(F^q_{\mu\nu})^2$ will be assumed to be fixed and static.

Structure of the paper: section \ref{sec2} presents the main results of our emulations for fields U(1), SU(2) and SU(3) with different tight-binding systems made of dimers and trimers. In section \ref{sec3} we discuss the construction of lattices with internal degrees of freedom using the energy levels of polymeric structures. Section \ref{sec4} is dedicated to the emulation of the U(1) abelian field using linear chains of rotating dimers. The emulation of the non-abelian field SU(2) is presented in section \ref{sec5}, where the phenomenon of wave confinement in a square lattice of dimers is achieved. The non-abelian vector field $\v A$ and the component $B_z$ of the field are plotted. Section \ref{sec6} contains some numerical results of the spectrum obtained from linear chains of rotating trimers.

\begin{figure}[b]
\centering
\includegraphics[width=0.3\textwidth]{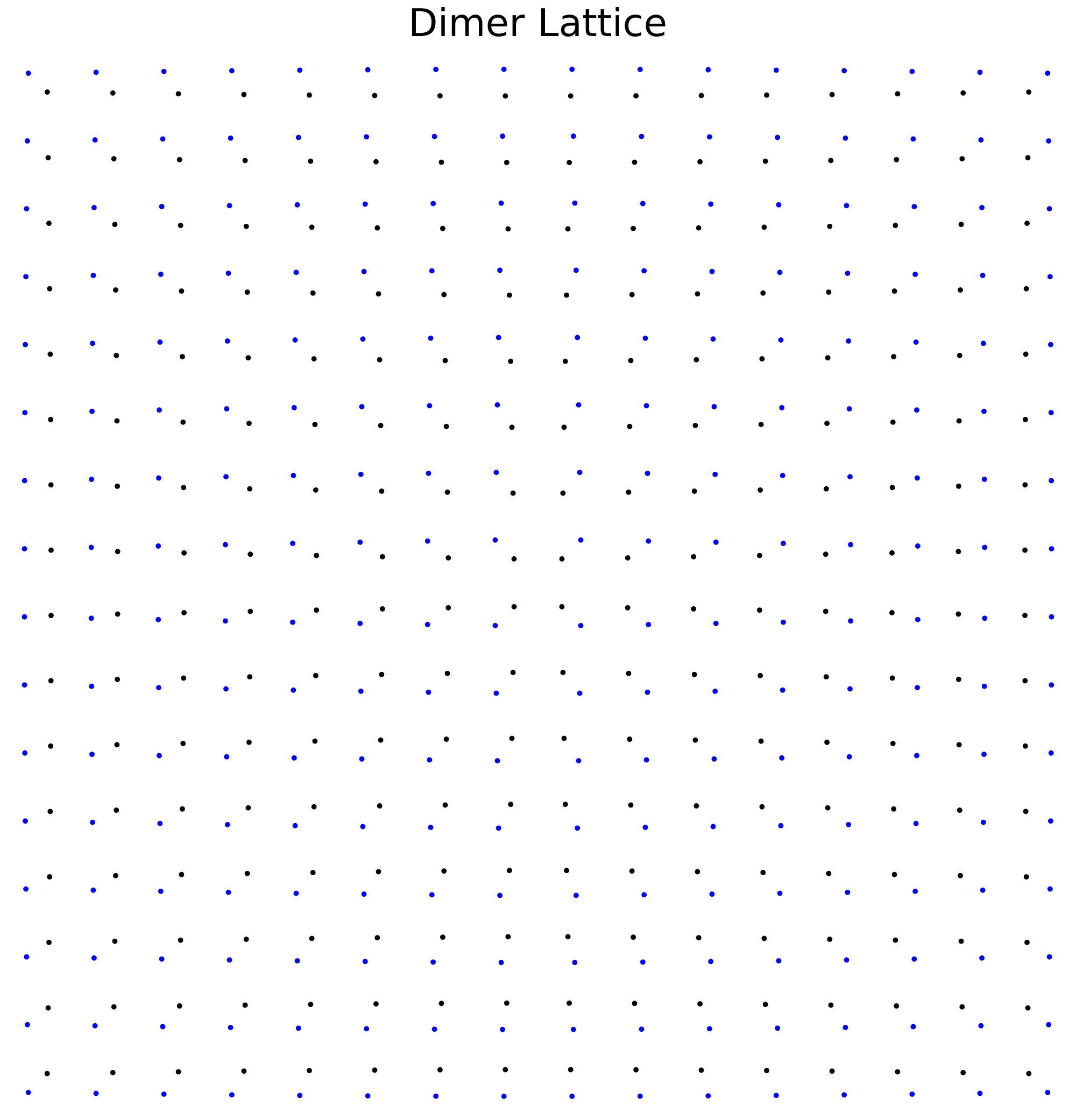}
\caption{\label{fig2} Hedgehog type lattice.  }
\end{figure}

\begin{figure}
\centering
\includegraphics[width=0.45\textwidth]{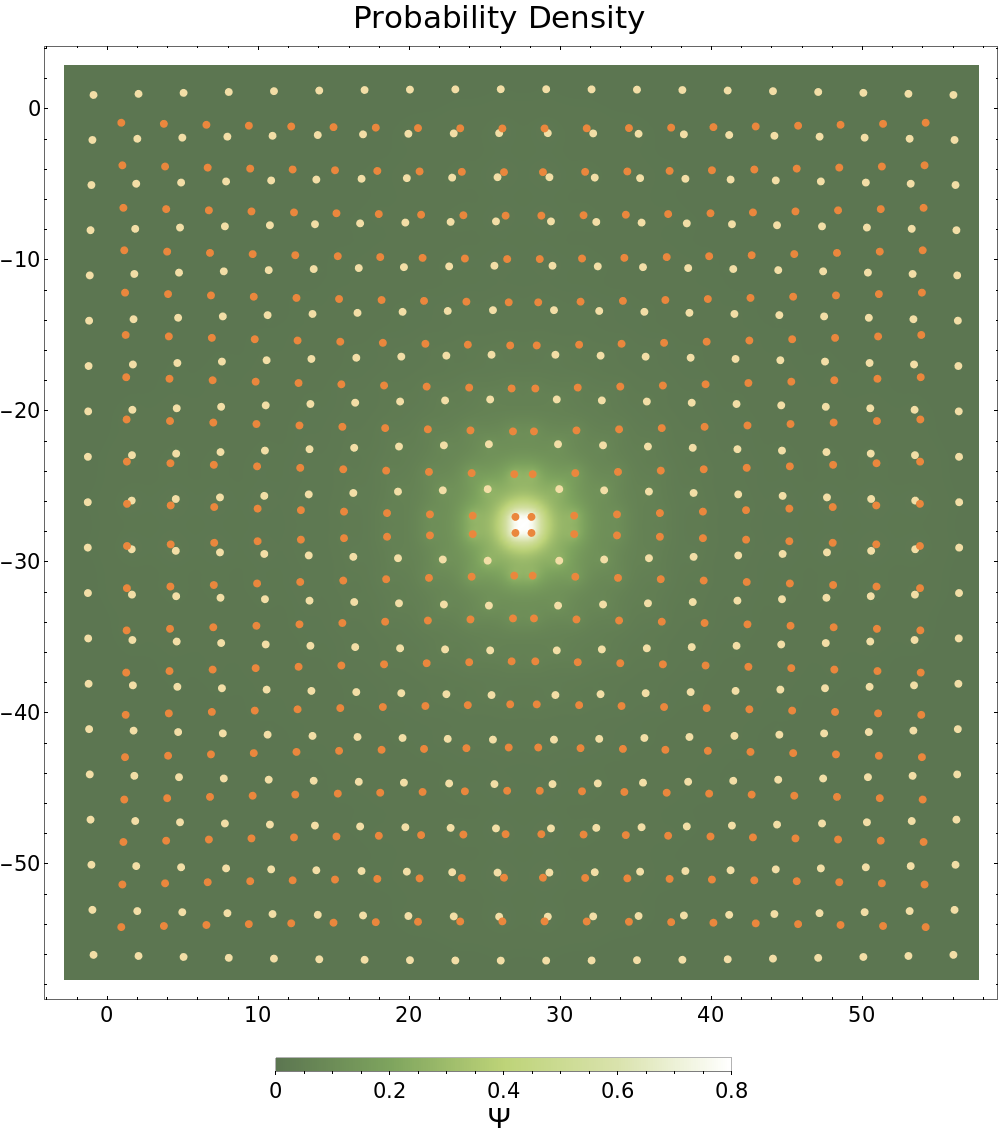}
\caption{\label{fig4} Square lattice with localized wave function at the center. The dimers are presented as a guide to the eye. The highest energy state displays the localization effect. The dimers are made of orange (dark) and yellow (light) dots, presented as a guide to the eye. The parameters are $L=2.9$, $d=1.3$ and $\lambda=1$.}
\end{figure}

\section{Presentation of results \label{sec2}}

In quantum field theory the magnetic field is considered  the starting point in the study of gauge invariance. Our first step is the emulation of the abelian U(1) theory. Historically, the effects of an external magnetic field acting on a two-dimensional lattice were studied by Peierls \cite{peierls1933} and later by  Harper\cite{harper1955}. The Harper model is a particular case of the Almost Mathieu Operator (AMO) \cite{last1995}, where the irrationality of the magnetic flux per unit cell  plays an important role in the shape of the spectral set. In 1964 \cite{azbel1964} Azbel found that in a lattice with an external magnetic field, each Landau level in the spectrum of the particle splits successively into sublevels. In 1976 Hofstadter found strong evidence of the fractal nature of the spectrum and subsequently it was proved that the spectral set must be a Cantor dust with zero Lebesgue measure \cite{hofstadter1976}. Emulations of the so-called Hofstadter butterfly have been proposed  in acoustic systems \cite{richoux2002,ni2019} and microwave scatterers \cite{kuhl1998}. 


{\sc {\rm U(1)} emulation with dimer chains.} We proceed with a construction of linear chains made of rotating dimers within a nearest-neighbor tight-binding description.  In this artificial realization, the number of complete turns $p$ of one dimer along the chain is the analog of the $\omega$ frequency --or magnetic flux-- in Harper's Hamiltonian. These chains exhibit singular spectra for specific lattice parameters, such as the distance between centers of neighboring dimers and the length of each dimer. Due to the internal degrees of freedom of each array site, the spectrum shows coupled bands. Two cases can be distinguished here: {\it i)} if the bands are strongly coupled, the non-abelian nature of the field emerges and {\it ii)} when these two bands are completely decoupled, we recover independent U(1) theories. The latter case is illustrated in Fig. \ref{fig1}.  The electronic structure can be studied properly if we write the Hamiltonian $H$ in the basis of symmetric and antisymmetric states of each dimer, and reorganize it into a block-diagonal form of couplings between symmetric levels (S--S) and antisymmetric levels (A--A). The spectrum generated by the S--S block is presented in panel (a) of Fig. \ref{fig1}, where the Hofstadter butterfly appears. This case emulates a magnetic field piercing through a square lattice and corresponds to $\Lambda=1$ in the AMO. On the other hand, panel (b) shows the spectrum of a magnetic field in a rectangular lattice, in this case $\Lambda$ is approximately 0.08$\cos\omega$. Both spectra show invariance when periodic boundary conditions are imposed, which is in agreement with the independence of the spectrum on Bloch's quasi-momentum $k$ along the periodic direction of a square lattice under the effect of an external magnetic field in the Landau gauge. It has been proved that the spectrum of the AMO is a Cantor set \cite{avila_martini}, so  in order to complete the analogy of our system with the Harper's model, we report the fractal dimensions of the antisymmetric and symmetric bands for fixed values of $\omega(N-1)/2\pi$ in Fig. \ref{fracdim} and Table \ref{tablefractal} respectively.

\begin{figure}[b]
\includegraphics[width=0.45\textwidth]{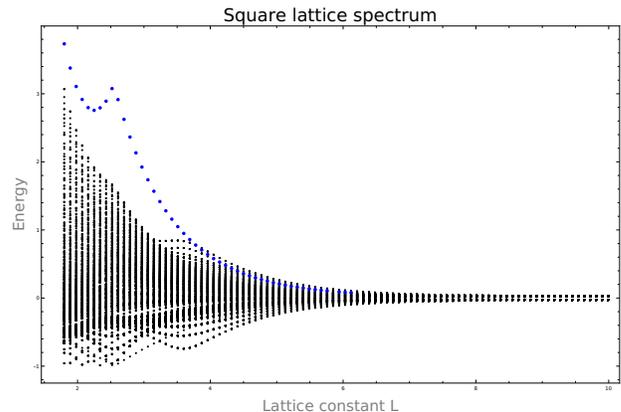}
\caption{\label{sqrlattice} Spectrum of the hedgehog type lattice as a function of lattice parameter $L$. Blue points represent  localized states at the center of the array as shown in Fig. \ref{fig4}. This state merges with one of the bands as the interdimer distance increases.}
\end{figure}

{\sc {\rm SU(2)} emulation and wave confinement.} The hypothesis of color confinement in quantum chromodynamics (QCD) establishes that only color-neutral multiplets of SU(3) can be observed. In the same spirit of the previous section, we construct a square lattice made of dimers in order to emulate the phenomenon of localization in SU(2). Although wave localization has been studied also in disordered arrays --Anderson localization-- we are interested here in stronger effects, e.g. bound states localized around a central point. The lattice is illustrated in Fig. \ref{fig2} and  the details of its construction will be discussed in section \ref{sec5}. Figure \ref{fig4} shows the localized wave function of the highest energy state on a lattice made of 400 dimers, whose sites are included as spatial reference.  We can observe that the confinement manifests around the region with the strongest couplings. This localized state coalesces with the propagation band and disappears in the limit when the lattice constant --distance between dimer centers-- is much greater than the length of the dimer \ie the fully periodic case without any internal structure.

\begin{figure}
\begin{tabular}{c}
\includegraphics[width=0.45\textwidth]{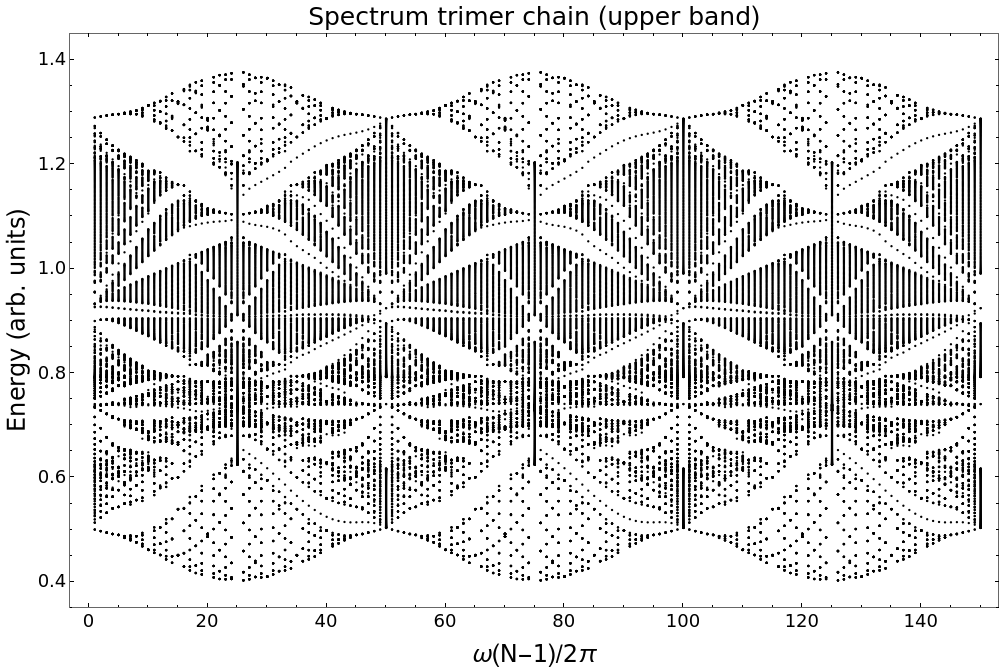}\\
\\
{\large (a)}\\
\\
\includegraphics[width=0.45\textwidth]{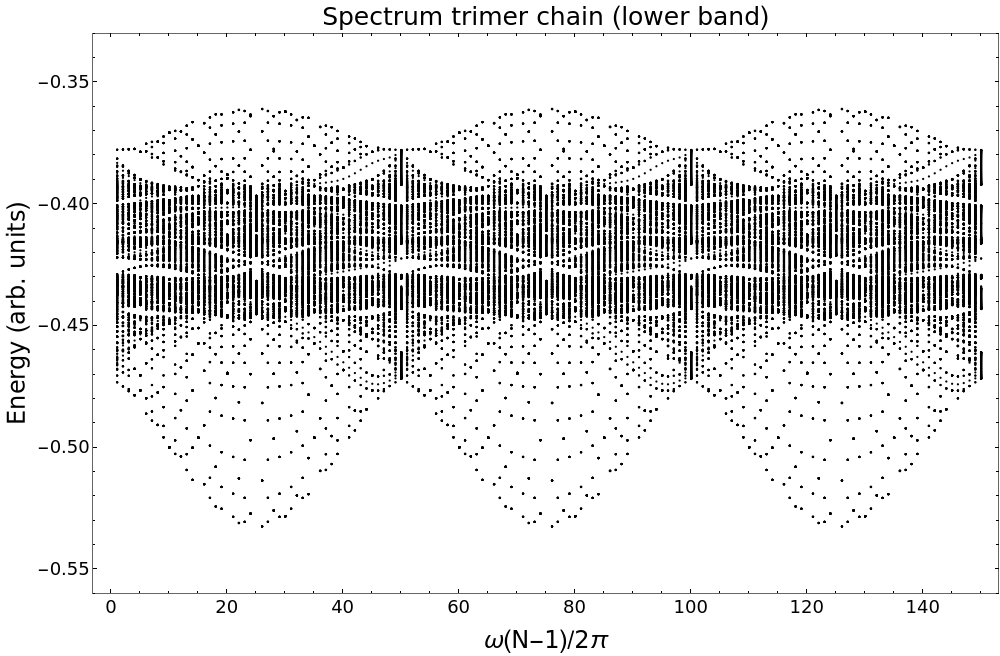}\\
\\
{\large (b)}\\
\\
\includegraphics[width=0.45\textwidth]{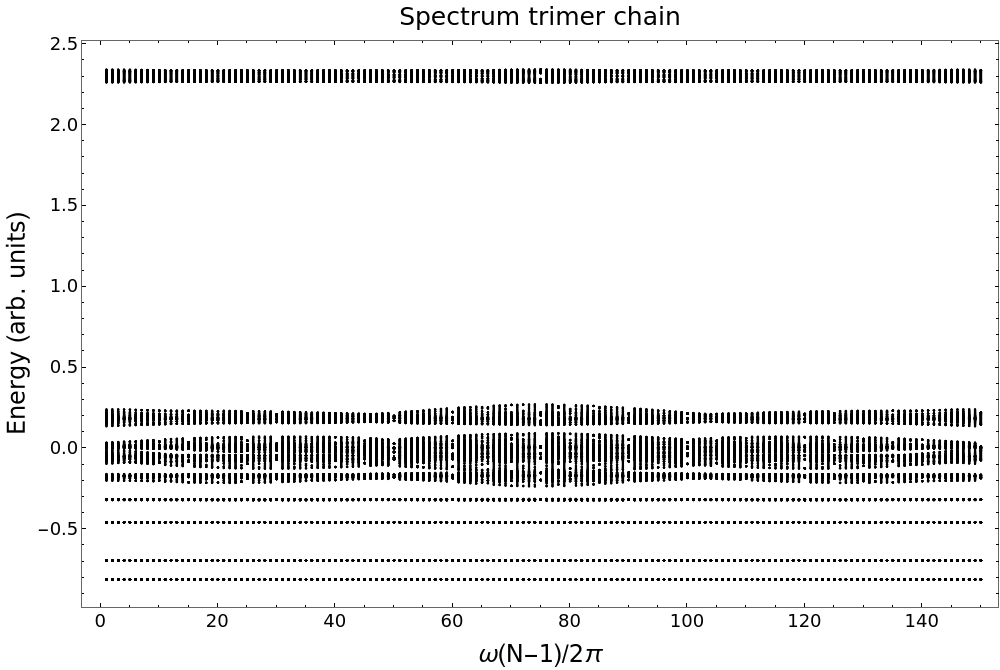}\\
\\
{\large (c)}
\end{tabular}
\caption{\label{figtrimers} The spectra of a linear chain made of trimers. The unit cell of the chain is illustrated in Fig. \ref{fig11}. a) The upper part of the spectrum is made of two overlapped bands. A fractal structure can be distinguished, resembling a Hofstadter's butterfly. b) Lower part of the spectrum. The band is dense and shows no recognizable structure. c) Spectrum with internal and global $C_3$ broken symmetry. The lower bands are flat.  }
\end{figure}

{\sc Linear chain with SU(3)}. Analogously to the linear chain of dimers, the SU(3) case can be studied by adding more degrees of freedom to each site. We propose a polymer of nine sites illustrated in Fig. \ref{fig11}. Some interesting results arise from this model. The spectrum of the system consists of three bands and two of them are overlapped as shown in Fig. \ref{figtrimers} panels a) and b). The upper band shows fractal structure consistent with Hofstadter's butterfly, while the lower band shows a dense structure with no recognizable pattern. The unitary cell of the chain consists of a trimer made of trimers as shown in Fig. \ref{fig11}. The separation of the bands can be performed by symmetry breaking of {\it i)} the global $C_3$ symmetry by rotation of only one internal trimer and {\it ii)} internal $C_3$ symmetry by deformation of the internal trimers. In panel c) of Fig. \ref{figtrimers} the lower part of the spectrum shows four decoupled bands with very small bandwidths. The study of superconductivity associated with flat bands might be of interest, as shown elsewhere\cite{deng_2003}. In our case, the spectrum shows localized states around specific energy values when the magnetic flux varies.

\begin{figure}
\includegraphics[width=0.45\textwidth]{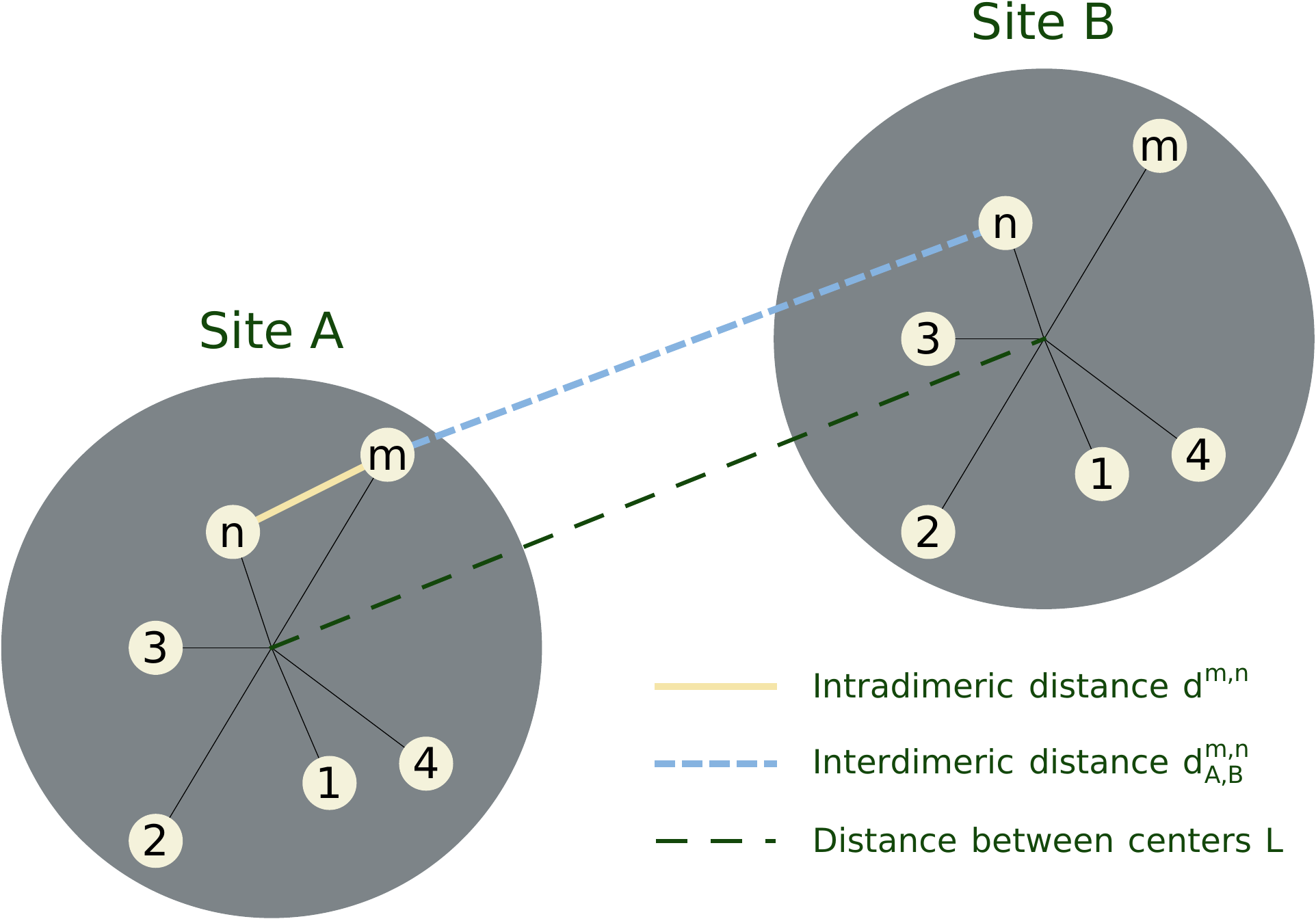}
\caption{Neighboring polymer sites. Two types of couplings can be distinguished here: {\it i}) intradimeric, which connects two sites of the same polymer (solid yellow line) and {\it ii}) interdimeric, which couples two sites from different polymers (blue dashed line). The distance between polymer centers (green dashed line) is assumed constant.  \label{polymer} }
\end{figure}

\begin{figure*}
\includegraphics[width=0.8\textwidth]{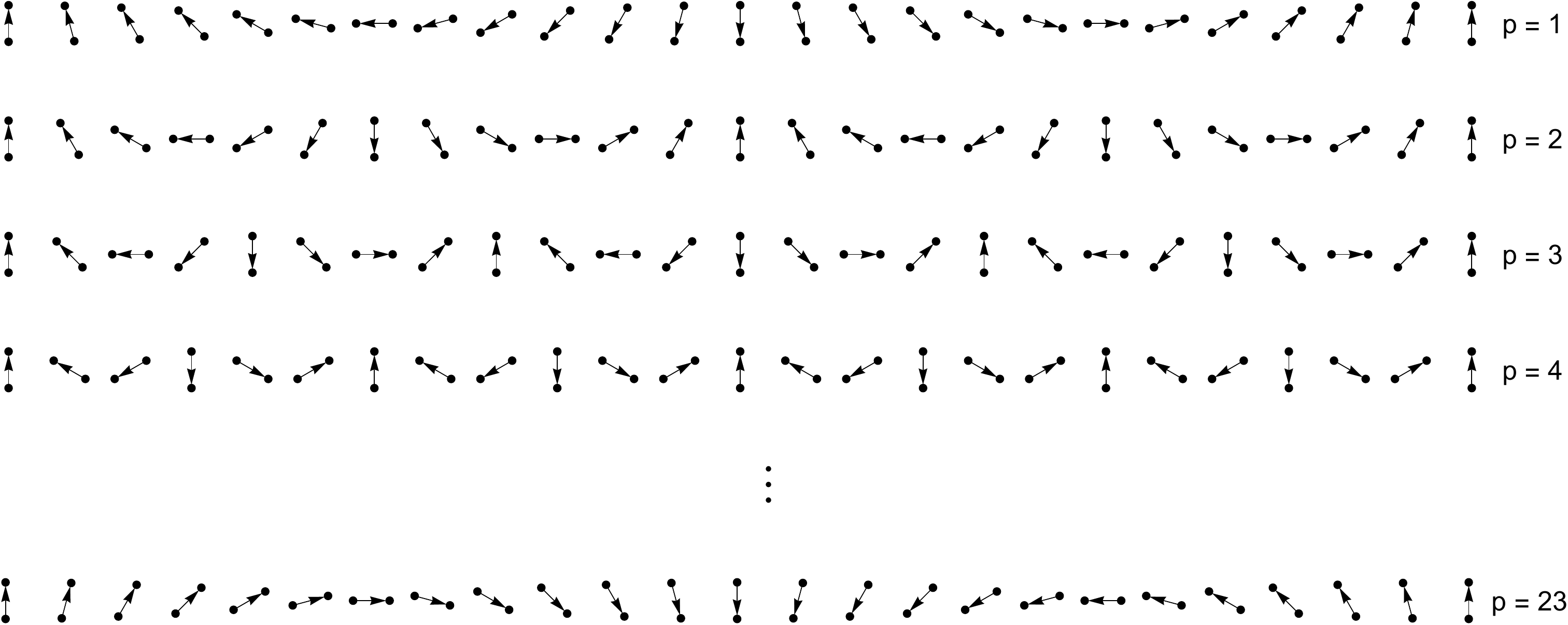}
\caption{Chains of rotating dimers. The arrows go from site 1 to site 2 of each dimer. The index $p$ stands for the number of complete rotations of one dimer along the chain. This parameter is analogous to the frequency $\omega$ in Harper's Hamiltonian as shown in eq. (\ref{eq:40}) and eq. (\ref{eq:41}).}
\label{fig:linearchain}
\end{figure*}

\section{Internal degrees of freedom: gauge group obtained from the lattice\label{sec3}}

The results presented in section \ref{sec2} provide evidence of a U(1) field emulation on two-dimensional lattices and the localization phenomenon in the SU(2) case. In order to properly formulate gauge fields in tight-binding arrays, the analog of minimal coupling in discrete systems must be introduced. This section is devoted to such formalism in discrete space with their correct laws of transformation, including non-abelian fields.

\subsection{General theory on the lattice\label{sec3.1}}

The usual treatment in continuous variables includes non-abelian fields $A_{\mu}=A_{\mu}^{\sigma}\tau_{\sigma}$, with $\mu$ a space-time index, $\sigma$ a group index and $\tau$ the generators of SU$(N)$. The minimal coupling prescription for space and time, ensures the gauge invariance of a theory involving charged particles and dynamical bosonic fields

\bea
p_{\mu} \mapsto p_{\mu} + g A_{\mu}, \qquad \v p \mapsto \v p + g\v  A.
\label{1.1}
\eea
This can be specialized to static fields where the 0-th component is irrelevant (purely magnetic); this is expressed in the second relation above and it is useful when considering static realizations in crystals. Upon the action of a unitary group U$(N)$, one has the transformations (here $\hbar=1$)

\bea
p_{\mu} \mapsto \tilde p_{\mu} = U^{\dagger} p_{\mu} U, \\
A_{\mu} \mapsto \tilde A_{\mu} = U A_{\mu} U^{\dagger} - \frac{i}{g} U \partial_{\mu} U^{\dagger}, \\
| \psi \> \mapsto U |\psi\>
\label{1.2}
\eea
and they are such that $p_{\mu} + g A_{\mu} \mapsto U^{\dagger}(p_{\mu} + g \tilde A_{\mu})U$ remains an invariant, \ie the transformation of $A$ compensates for the unitary transformation of the operators and physical states. In our discussions, this standard formulation has to be deduced in the spirit of (\ref{1.1}): A stationary Hamiltonian formulation in first quantization necessitates a Schr\"odinger-like equation coming from $p_0$, so a gauge invariant spectral problem involves now the transformation

\bea
\tilde H = U^{\dagger} H U, \qquad \partial_0 U = 0
\label{1.3}
\eea
and a static field such that

\bea
A_0 = {\rm constant}, \quad  \tilde{\v A} = U \v A U^{\dagger} - \frac{i}{g} U \v \nabla U^{\dagger}.
\label{1.4}
\eea
When it comes to derivatives on the lattice, we have to consider finite-difference operators. The aim is to identify the functional dependence of $H$ on the field $A$ using the correct form of minimal coupling in discrete variables. For Hamiltonians that depend of translation operators $T_i$ and lattice vectors $\v R$, we use the notation

\bea
&\v R = \sum_{i=1}^{m} n_i \v a_i \\
&N_i |n_1,...,n_m \> = n_i |n_1,...,n_m \>,\\
&T_i |n_1,...,n_i,...,n_m \> = |n_1,...,n_i +1,...,n_m \>.
\label{1.5}
\eea
where $\v a_i$ stand for primitive vectors and $m$ for the number of such vectors in the array. In our realizations, the lattice states $|n_1,...,n_m\>$ are also represented by localized wave functions around sites --e.g. a trapped mode inside a cylindrical dielectric cavity. In this language, the operators $T$ have a differential form

\bea
\< \v r | T_i |n_1,...,n_2 \> &=& \< \v r |n_1,...,n_i+1,...,n_2 \> \nonumber \\
&=& \< \v r -\v a_i |n_1,...,n_m \> \nonumber\\
&=& \exp(-i \v a_i \cdot \v p) \< \v r |n_1,...,n_m \>.
\label{1.6}
\eea
 It is not difficult to guess that minimal couplings should be given by complex exponentials in front of $T$: Indeed, under a static gauge transformation 

\bea
\tilde T_i = \exp(-i \v a_i \cdot  U^{\dagger} \v p U ) = U^{\dagger} T_i U.
\label{1.7}
\eea
Moreover, a local dependence of the transformation $U(N_1,...,N_m) \equiv U(\v N)$ allows to write

\bea
\tilde T_i = U^{\dagger}(N_1,...,N_2) U(N_1,...,N_i-1,...,N_m) T_i, 
\label{1.8}
\eea
or $\tilde T_i = U^{\dagger}(\v N) U(\v N - \v e_i) T_i $, which are the transformations we want to emulate. In order to compensate for this transformation, one writes space-dependent couplings in front of translation operator as

\bea
H = \sum_{k, i} \Delta_{k,i}(\v N) (T_i)^k + {\rm h.c.} + V(\v N),
\label{1.9}
\eea
where $V$ is a local potential, $k$ is the range of the interaction, and $\Delta_{k,i}$ are the coupling constants as functions of site operators; in our case, they are real and positive. The aforementioned invariance imposes a transformation of $\Delta$ of the type

\bea
\tilde \Delta_{k,i} (\v N) = \Delta_{k,i} (\v N) U^{\dagger}(\v N - \v e_i) U(\v N),
\label{1.10}
\eea
thus compensating the extra factors appearing in (\ref{1.8}). If we want to know the specific dependence of $H$ on $A$, the abelian case yields simply

\bea
\label{1.11}
\Delta_{k,i}(\v N) = | \Delta_{k,i}(\v N) | \times \exp\{- i g \, \v a_i \cdot \v A(\v N)\},\\
 | \Delta_{k,i}(\v N) | \equiv | \Delta_{k,i}(\v N) |^{\dagger}.
\eea
Here, the hermitian part of $\Delta$ has been written explicitly to show that the field appears inside a unitary operator, while the {\it modulus\ } is still allowed to depend on $\v N$ due to lattice deformations, such as those considered in \cite{rivera-mocinos_inverse_2016,franco-villafane_first_2013,sadurni_playing_2010}. A gauge transformation can be performed now at the level of $A$; by defining a set of orthogonal vectors --not necessarily on the physical lattice-- $\hat b_1 \perp \v a_1, \hat b_2 \perp \v a_2$, the displacement by a discretized derivative

\begin{widetext}
\bea
\v A(\v N) \mapsto \v A(\v N) + \frac{\hat b_1}{\hat b_1 \cdot \v a_2} \left[ \Phi(\v N - \v e_1)- \Phi(\v N) \right] + \frac{\hat b_2}{\hat b_2 \cdot \v a_1} \left[ \Phi(\v N - \v e_2)- \Phi(\v N) \right]
\label{1.12}
\eea
\end{widetext}
produces exactly the transformation (\ref{1.8}) when substituted in (\ref{1.11}). The unitary operator in this case is $U(\v N)= e^{ig \Phi(\v N)}$.

For non-abelian fields, one does not express $\Delta$ directly as an exponential of $A$. Instead, one has to solve a relation of the type

\bea
U^{\dagger}(\v N) \Delta_{k,i} \left[ \tilde{\v A} \right] (T_i)^k U(\v N) = \Delta_{k,i} \left[ \v A \right] (T_i)^k
\label{1.13}
\eea
which must compensate once more (\ref{1.8}), so now

\bea
U^{\dagger}(\v N) \Delta_{k,i} \left[ \tilde{\v A} \right] U(\v N) = \Delta_{k,i} \left[ \v A \right] U^{\dagger}(\v N - \v e_i) U(\v N)\nonumber\\
\label{1.14}
\eea
and since 

\bea
\tilde{A_j} = U(\v N) \left[ A_j - \frac{i}{g}(1- U^{\dagger}(\v N - \v e_j)U(\v N)) \right]U^{\dagger}(\v N)\nonumber\\ 
\label{1.15}
\eea
one arrives to the functional relation

\bea
&\Delta_{k,i} \left[ A_j - \frac{i}{g}(1- \exp \left[ -i g \, \Phi(\v N - \v e_j) \right]\exp \left[ -i g \, \Phi(\v N) \right]) \right]\nonumber\\ 
&= \Delta_{k,i} \left[ A_j \right] \times \exp \left[ -i g \, \Phi(\v N - \v e_i) \right]
\label{1.16}
\eea
where $\Phi = \Phi^{\sigma}\tau_{\sigma}$ is now an arbitrary non-abelian  operator. This functional relation determines $\Delta$ in terms of $A$ recursively, but its explicit solution is difficult to obtain in general. Instead we say that there is a non-trivial gauge field acting on a particle described by Hamiltonian $H$ if there is no transformation $U$ for which

\bea
\Delta_{k,i}\left[ 0 \right] U(\v N - \v e_i)= U(\v N) \Delta_{k,i} \left[ \v A \right]. 
\label{1.17}
\eea 
Even in the abelian case, this demands the use of non-trivial structures such as complex hopping amplitudes and, as a consequence, singular spectra via Harper's equation. 

We have achieved the introduction of minimal coupling and gauge theory in the context of tight-binding systems with single-level sites, e.g. arrays of resonators. The U(1) field will be implemented in section  \ref{sec4.1}  with the construction of a linear chain made of dimers.  This system possesses two internal degrees of freedom and the couplings $\Delta$ are modulated by the rotation of each dimer along the chain. We will show that in the appropriate basis of symmetric and antisymmetric dimeric states, two decoupled blocks emerge, both emulating singular spectra similar to those obtained from Harper's Hamiltonian. The phenomenon of wave confinement with the SU(2) field is presented in section \ref{sec5}.

\begin{figure*}\centering
\includegraphics[width=13cm]{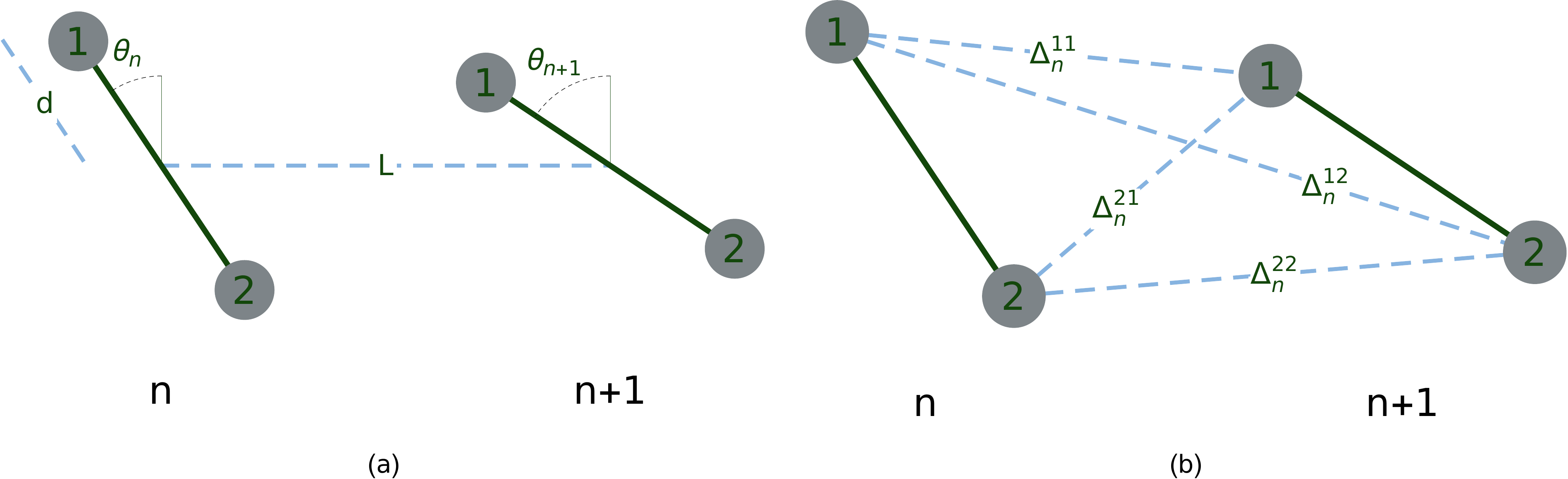}
\caption{(a) Geometric parameters for two adjacent dimers. (b) All pairwise inter dimer couplings. Each dimer is rotated by an angle $\theta_n$ which is a function of the number $N$ of dimers in the chain and the number $p$ of complete rotations of one dimer along the chain. (b) All possible pairwise couplings between sites in the same array of dimers as in (a). The couplings depend on parameters $d$, $L$ and $\theta_n$.}
\label{pairwisecoup}
\end{figure*}

\subsection{Decomposition in terms of external lattice couplings and internal polymeric couplings\label{sec3.2}}

The general idea to include an internal SU$(N)$ group, consists in the introduction of internal levels corresponding to each lattice site. Since one site corresponds to a single level resonator, we consider a set or cluster of such resonators put tightly together, resembling a polymer. Then, each of these sets, when regarded as a single site with a complex structure, shall have as many levels as sites in the polymer. See Fig. \ref{polymer}. 
In connection with couplings, it is also reasonable to decompose them into two parts using a diagonalization trick acting exclusively on each polymer. This helps to keep track of their contribution to the global lattice defined by the location of each polymer, or its geometric center. The distance between centers shall be associated with a coupling $\Delta$ referred to as external coupling. The distance between a geometric center and all other sites of the same polymer shall be regarded as an internal distance, therefore we refer to the corresponding $\Delta$ as internal coupling. Some triangular inequalities can be established for such couplings, as long as they are given by monotonically decreasing functions of the separation distances, e.g. exponential functions as in \cite{kuhl2010,barkhofen2013,bellec2013} or even more careful approximations involving modified Bessel functions. Based on geometric relations for the distances, we can say that the internal couplings in each polymer are more important in magnitude than the external ones. However, the external couplings shall reflect the desired crystalline structure, while internal bonds shall give rise to a local non-abelian field acting as a perturbation. Such a perturbation can be either weak or strong, depending on the geometry.  Once this decomposition is achieved, we will be ready to use an appropriate basis to describe the polymeric states that interact with other neighboring polymeric states. In fact, this has a close resemblance with standard tight-binding models applied to solid state physics, where individual sites have more than one conduction orbital \cite{sutton1993}. The diagonalization of isolated polymers does the job, eliminating thus the internal couplings between polymeric eigenstates and revealing new site-to-site interactions.

As it is evident, the number of sites in a polymer is equivalent to the dimensionality of the internal group. We note that the resulting theory is abelian, if the polymeric states are coupled only to their neighboring counterparts with the same index (\ie first level with first level, second level with second level and so forth). This gives rise to a number of $N$ decoupled abelian theories on the same lattice. On the other hand, if the polymeric states are connected with different internal eigenstate indices of a neighboring site, the corresponding Hamiltonian shall contain non-diagonal operators defined on the generating algebra su$(N)$. This shall give rise to a non-abelian theory, where the external field appears in the coupling constants instead of a local potential; this is precisely what we showed in (\ref{1.17}).

We use the following notation: $\sigma_{\pm}$ is the operator that rises (lowers) site number inside a polymer, $T_i$ is the translation operator between neighbors in the direction of a primitive vector $\v a_i$, $d$ is the distance between the center of a polymer and one of its sites (for simplicity, we take all these as equal), $L$ is the distance between geometric centers of neighboring polymers, and $\Delta(d_{A,B}^{m,n})=\Delta_{A,B}^{m,n}$ is the coupling as a function of the distance as Fig. \ref{polymer} shows.


\begin{figure}[b]
\begin{tabular}{c}
\includegraphics[width=0.45\textwidth]{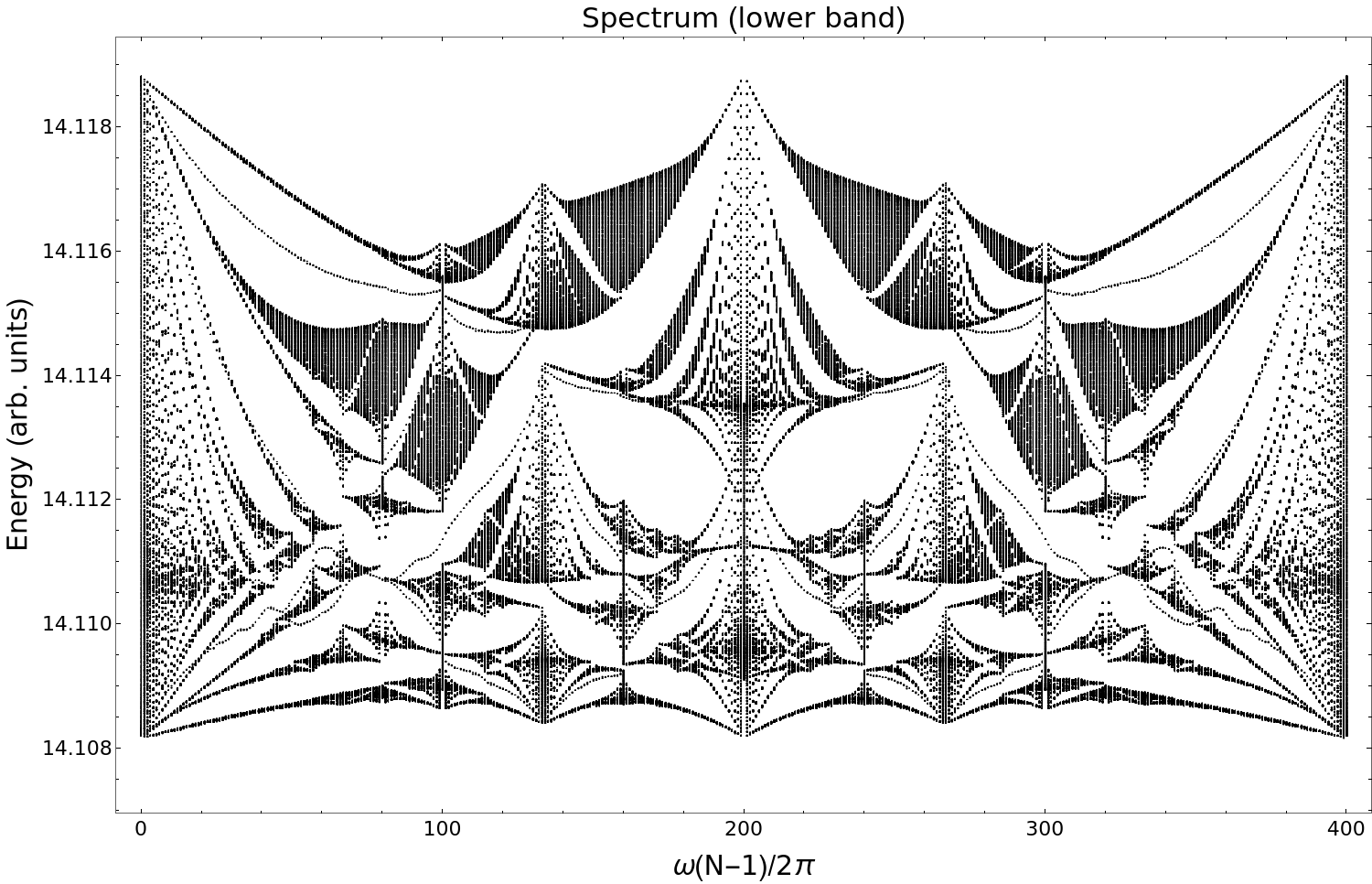}\\
\\
{\large (a)}\\
\\
\includegraphics[width=0.45\textwidth]{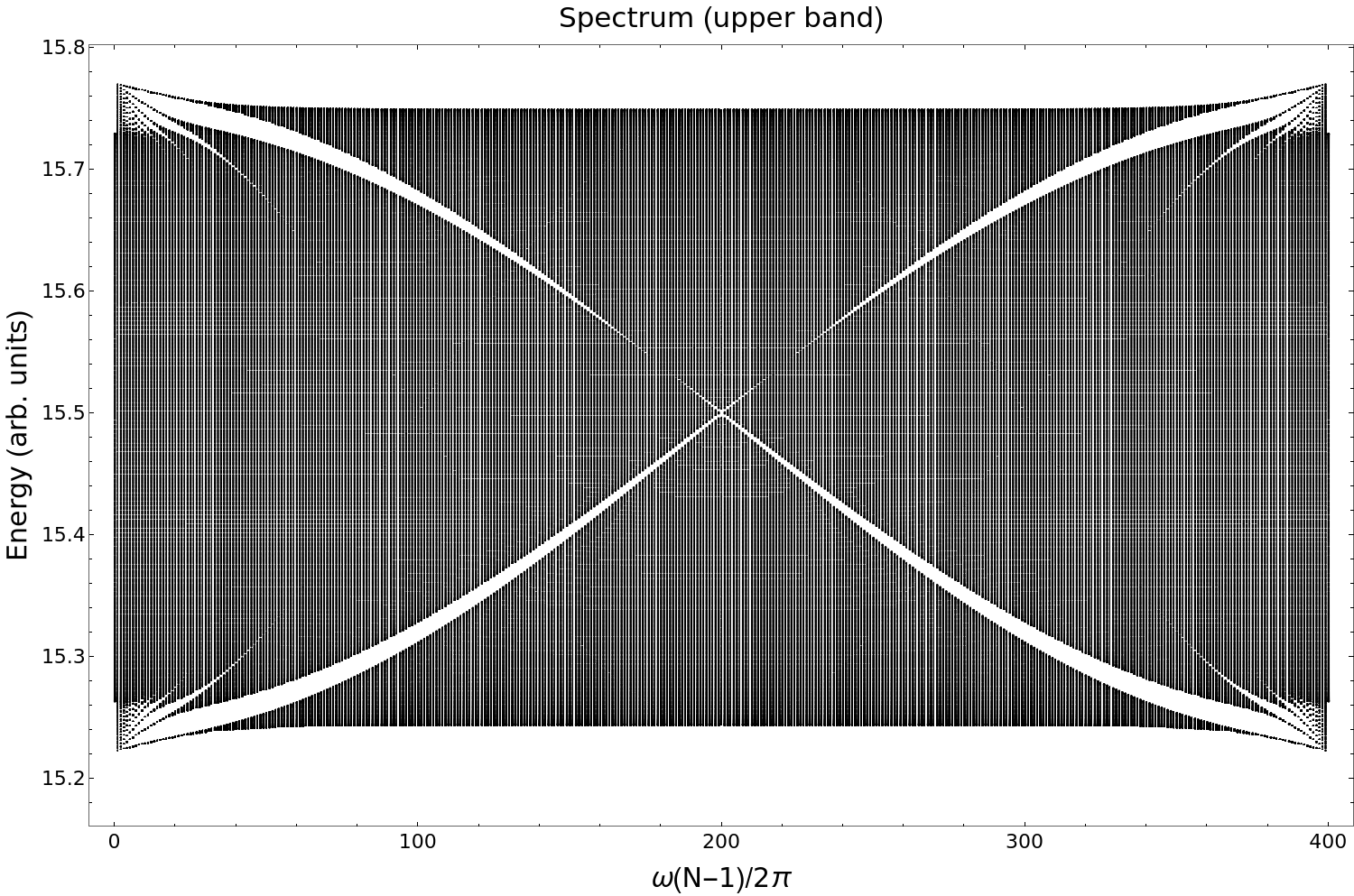}\\
\\
{\large (b)}
\end{tabular}
\caption{\label{fig9} Deformed fractal spectrum due to off-diagonal couplings produced by an SU(2) field. (a) Strong corrections to the symmetric mode band. (b) Mild corrections to the antisymmetric mode band. }
\end{figure}

\begin{figure*}
\includegraphics[width=0.7\textwidth]{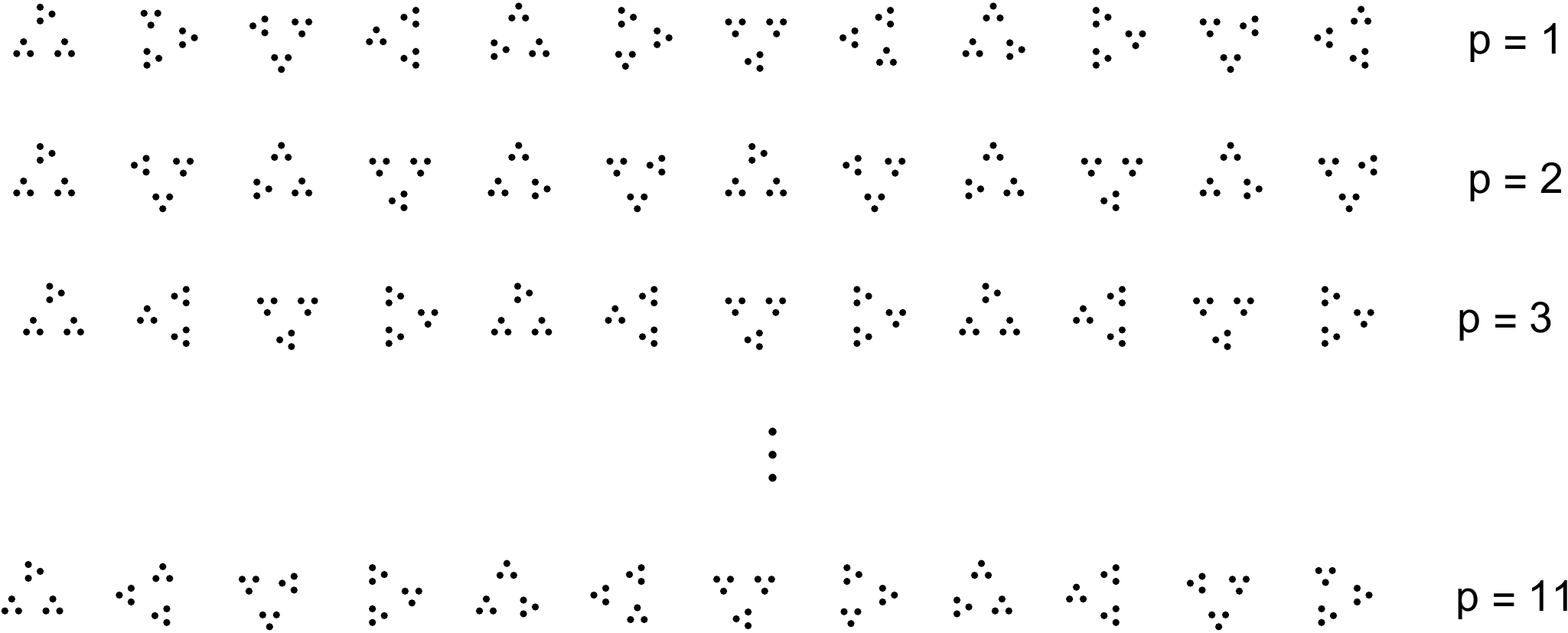}
\caption{Linear chains made of trimers. The unit cell rotates a number $p$ of complete turns along the chain. The global $C_3$ symmetry is broken if one internal trimer is rotated with respect to the other two. When the symmetry $C_3$ is restored a flat band appears in the inferior part of the spectrum.  This symmetry breaking splits the resulting energy bands. }
\end{figure*}

\section{Emulation of Peierls substitution for U(1) gauge field\label{sec4}}


\begin{figure*}[t]
\centering
\begin{tabular}{cc}
\includegraphics[width=0.40\textwidth]{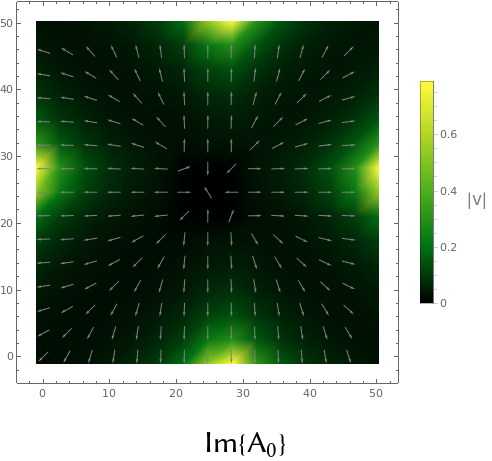}&
\includegraphics[width=0.40\textwidth]{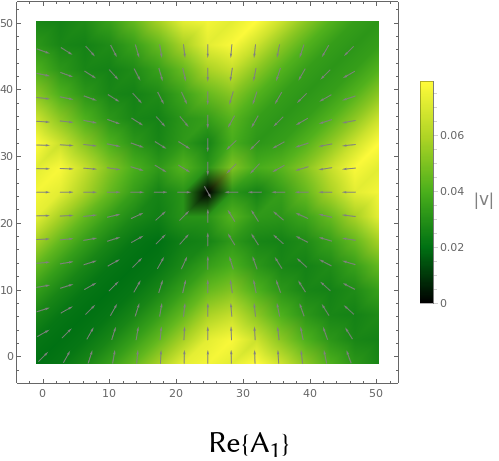}\\ \\ \\
\includegraphics[width=0.40\textwidth]{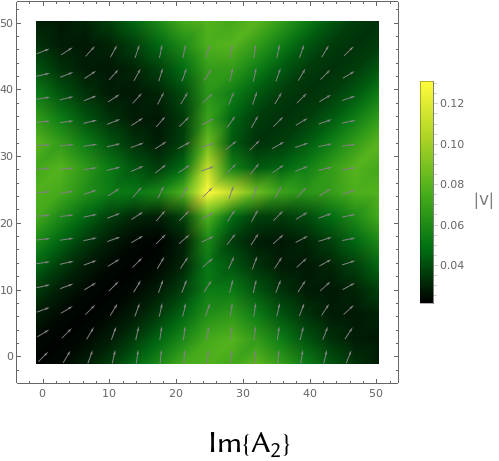}&
\includegraphics[width=0.40\textwidth]{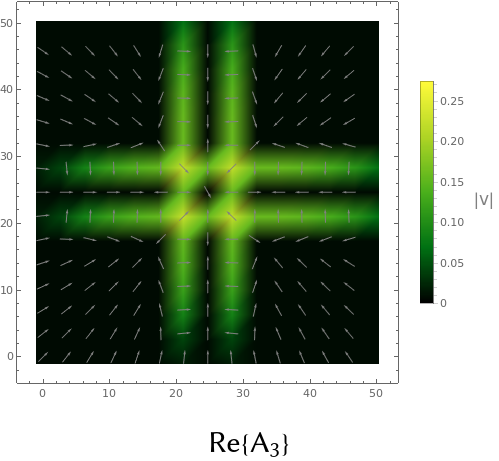}\\
\end{tabular}
\caption{\label{afield1} Portraits of $\v A$ on the square lattice. The components Im$\{A_0\}$ and Im$\{A_2\}$ are presented, as  Re$\{A_0\}=$Re$\{A_2\}=0$. Similarly the components Re$\{A_1\}$ and Re$\{A_3\}$ are presented, while  Im$\{A_1\}=$Im$\{A_3\}=0$.}
\end{figure*}

\subsection{Linear chain\label{sec4.1}}

As announced in section \ref{sec2}, we introduce the construction of dimeric chains, as can be visualized in Fig. \ref{fig:linearchain}. Each chain is characterized by the parameter $p$ that corresponds to the number of complete turns of one dimer along the array. The localized states around each site are denoted by $|n,i\>$ where $n$ stands for the dimer position and $i$ is the intradimeric site. The couplings are given by: 
\bea\label{eq23}
\Delta_{n,m}^{ij}&=&\<n,i |H|m,j \>\nonumber\\
&=&\int  \psi_i^* (r)H(-\nabla ^2,\v r)\psi_j(r) dr \nonumber\\
&\sim& \Delta_0\exp(-d_{n,m}^{i,j}/\lambda).
\eea
where $d_{n,m}^{i,j}$ is the distance between sites (n,i) and (m,j) and $\lambda$ is the evanescence length. In our description only nearest neighbors are considered. The notation for the couplings is simplified as follows: $\Delta_{n,n+1}^{ij}\mapsto \Delta_{n}^{i,j}$. We write the Hamiltonian of one chain as:   
\be
H_p=\left(
\begin{array}{cccccc}
H_0&h_1&0&\cdots&&0\\
h_1^\dagger&H_0&h_2&\ddots&&0\\
0&h_2^\dagger&H_0&&&0\\
\vdots&\ddots&&\ddots&&0\\
&&&&H_0&h_{n-1}\\
0&&&0&h_{n-1}^\dagger&H_0
\end{array}
\right),
\ee
where 
\be
H_0=
\left(
\begin{array}{cc}
0&\Delta_0\\
\Delta_0&0
\end{array}
\right),\quad
h_n=
\left(
\begin{array}{cc}
\Delta_n^{11}&\Delta_n^{12}\\
\Delta_n^{21}&\Delta_n^{22}
\end{array}
\right).
\ee
In this notation, $H_0$ represents the Hamiltonian of one dimer with its on-site energies shifted to zero for simplicity. Now we can diagonalize each dimeric block in the main diagonal using a Hadamard matrix:
\be
\small
H_p^{(1)} = \left(\mathbf{1}_N \otimes U \right)^\dagger H_p \left(\mathbf{1}_N \otimes U \right), \quad
U=\frac{1}{\sqrt{2}}
\left(
\begin{array}{ll}
+1&+1\\
+1&-1\\
\end{array}
\right).
\ee
The transformed Hamiltonian in the symmetric and antisymmetric basis is:  
\be
H_p^{(1)}=\left(
\begin{array}{cccccc}
H_D&\tilde{h}_1&0&\cdots&&0\\
\tilde{h}_1^\dagger&H_D&\tilde{h}_2&\ddots&&0\\
0&\tilde{h}_2^\dagger&H_D&&&0\\
\vdots&\ddots&&\ddots&&0\\
&&&&H_D&\tilde{h}_{n-1}\\
0&&&0&\tilde{h}_{n-1}^\dagger&H_D
\end{array}
\right),
\ee
with $H_D$ and $\tilde{h}_i$ given by:
\be
H_D=
\left(
\begin{array}{cc}
\Delta_0&0\\
0&-\Delta_0
\end{array}
\right),\quad
\tilde{h}_n=
\left(
\begin{array}{cc}
\Delta_n^{AA}&\Delta_n^{AS}\\
\Delta_n^{SA}&\Delta_n^{SS}
\end{array}
\right).
\ee
The superscript of $\Delta_n$ in $\tilde{h}_n$ now represents the couplings between antisymmetric (A) and symmetric (S) states. We express $\Delta_n^{AA},\Delta_n^{SA},\Delta_n^{AS}$ and $\Delta_n^{SS}$ in terms of site-to-site couplings:
\bea\label{eq29}
\Delta_n^{AA}=\frac{1}{2}\left(\Delta_n^{11}+\Delta_n^{12}+\Delta_n^{21}+\Delta_n^{22}  \right)\nonumber\\
\Delta_n^{AS}=\frac{1}{2}\left(\Delta_n^{11}-\Delta_n^{12}+\Delta_n^{21}-\Delta_n^{22}  \right)\nonumber\\
\Delta_n^{SA}=\frac{1}{2}\left(\Delta_n^{11}+\Delta_n^{12}-\Delta_n^{21}-\Delta_n^{22}  \right)\nonumber\\
\Delta_n^{SS}=\frac{1}{2}\left(\Delta_n^{11}-\Delta_n^{12}-\Delta_n^{21}+\Delta_n^{22}  \right).
\eea
Reorganization of the transformed Hamiltonian $H_p^{(1)}$ into A and S blocks obtains 
\be\label{eq30}
H_p^{(2)}=
\left(
\begin{array}{cc}
H_A&\delta_{AS}\\
\delta_{AS}^\dagger &H_S
\end{array}
\right),
\ee
where $H_A$ and $H_S$ contain only A--A and S--S couplings respectively: 
\bea
H_A=
\left(
\begin{array}{cccccc}
\Delta_0&\Delta_1^{AA}&0&\cdots&&0\\
\Delta_1^{AA}&\Delta_0&\Delta_2^{AA}&\ddots&&\vdots\\
0&\Delta_2^{AA}&\Delta_0&&&\\
\vdots&\ddots&&\ddots&\\
&&&&\Delta_0&\Delta_{n-1}^{AA}\\
0&&&&\Delta_{n-1}^{AA}&\Delta_0
\end{array}
\right)\\
H_S=
\left(
\begin{array}{cccccc}
\Delta_0&\Delta_1^{SS}&0&\cdots&&0\\
\Delta_1^{SS}&\Delta_0&\Delta_2^{SS}&\ddots&&\vdots\\
0&\Delta_2^{SS}&\Delta_0&&&\\
\vdots&\ddots&&\ddots&\\
&&&&\Delta_0&\Delta_{n-1}^{SS}\\
0&&&&\Delta_{n-1}^{SS}&\Delta_0
\end{array}
\right)\\
\delta_{AS}=
\left(
\begin{array}{cccccc}
0&\Delta_1^{AS}&0&\cdots&&0\\
\Delta_1^{SA}&0&\Delta_2^{AS}&\ddots&&\vdots\\
0&\Delta_2^{SA}&_0&&&\\
\vdots&\ddots&&\ddots&\\
&&&&0&\Delta_{n-1}^{AS}\\
0&&&&\Delta_{n-1}^{SA}&0
\end{array}
\right)
\eea 
We observe from (\ref{eq29}) that the elements of $\delta_{AS}$ depend on the couplings between neighboring sites, and this dependence is purely geometrical due to the definition of $\Delta_{n,m}^{ij}$ in (\ref{eq23}). The ideal case was announced in section \ref{sec2} where $\delta_{AS}$ is identically zero; the corresponding spectrum is illustrated in Fig. \ref{fig1}.  

In general, the Hamiltonian $H_p^{(2)}$ possesses two coupled blocks, although some approximations can be made. If the elements of $\delta_{AS}$ are smaller than the ones of $H_A$ and $H_S$, the spectrum consists of decoupled bands with a Cantor spectrum; in the more general case, the Hofstadter butterfly appears slightly deformed, as the reader may visualize in Fig. \ref{fig9}.

In order to obtain the couplings as functions of site number, we must calculate the four distances between the dimer $n$ and the dimer $n+1$. Let us denote by $\mathbf{r}_n^{(i)}, \, i=1,2$ the position vector of site $i$ in dimer $n$, so all the pairwise distances are given by the following expression:

\be
d_n^{ij}=|\v r_n^{(i)}-\v r_{n+1}^{(j)}|=L\left\{1+2\alpha^2 A_n^{ij} -2\alpha B_n^{ij}\right\}^{1/2},
\ee
where  $A_n^{ij}$ and $B_n^{ij}$ contain information on the rotations of each dimer $\theta_n=2\pi P(n-1)/(N-1)$ as

\bea
A_n^{ij}&=&1-(-1)^{i+j}\cos(\theta_n -\theta_{n+1})\nonumber\\
B_n^{ij}&=&(-1)^i \sin\theta_n - (-1)^j\sin\theta_{n+1},
\eea
and $\alpha=d/L$. The corresponding couplings are given as follows:

\be
\Delta_n^{ij}\simeq \left\{ 1+B_n^{ij}\frac{d\alpha}{\lambda}+\left[\frac{B_n^{ij^2}}{2}\left(1+\frac{L}{\lambda}\right)-A_n^{ij}   \right]\frac{d\alpha^2}{\lambda}\right\}\Delta_L.
\ee
where a second order approximation for $\alpha \ll 1$ has been used and $\Delta_L= \Delta_0 \exp(-L/\lambda)$. From (\ref{eq29}) the  second order approximation in $\alpha$ leads to
\be
\Delta_n^{SS}=\frac{2 d^2}{\lambda L}\left\{\cos\theta_n\cos\theta_{n+1}-\frac{L}{\lambda}\sin\theta_n\theta_{n+1}\right\}\Delta_L.
\ee
In the special case when $L/\lambda=1$, we obtain
\be\label{eq:40}
\Delta_n^{SS}=\frac{2d^2}{L^2}\cos\left[\omega(2n-1) \right]\Delta_L,
\ee
with the angular frequency $\omega$ defined as $2\pi p/(N-1)$. It can be demonstrated that there is a unitary map between the Harper's Hamiltonian and a tight-binding model without potential and with site-dependent couplings, as obtained in (\ref{eq:40}). This rigorous map is of utmost importance, and it is included in  Appendix \ref{appendixa}. The effective parameter is $\Lambda=1$ for the symmetric mode band. On the other hand, for $\Delta_n^{AA}$ couplings we have:  
\be\label{eq:41}
\Delta_n^{AA}=2\gamma\left\{1 -\frac{\beta\alpha^2(\beta+1)}{\gamma}\cos\omega\cos[\omega(2n-1)] \right\}\Delta_L.
\ee
where
\bea
\gamma&=&2+\beta\alpha^2(\beta-1),\nonumber\\
\beta &=&\frac{d}{\lambda},\quad \alpha=\frac{L}{d}.
\eea
In this case, $\Lambda$ in the Harper's model must be assumed small as shown in (\ref{eq:a16}). The estimated effective is $\Lambda=0.08\cos\omega$ for the antisymmetric mode band. It is worth noting the non-removable frequency dependence of $\Lambda$. The Lebesgue measure for the AMO is $|4-4\Lambda|$ \cite{avila_2006}. The symmetric and antisymmetric bands have zero and 3.68$|\cos\omega|$ Lebesgue measure respectively. 

A more general treatment to find the couplings as functions of site numbers, consists in a decomposition of $\tilde{h}_n$ in polar form:

\bea
\tilde{h}_n&=&UP=\exp(i\sigma_2 \phi)P\nonumber\\
&=&
\left(
\begin{array}{cc}
\cos\phi&\sin\phi\\
-\sin\phi&\cos\phi
\end{array}
\right)
\left(
\begin{array}{cc}
e^\theta \cosh\chi&\sinh\chi\\
\sinh\chi&e^{-\theta}\cosh\chi,
\end{array}
\right)
\eea  
obtaining thus the following relations
\bea
\Delta_n^{AA}&=&e^\theta\cos\phi\cosh\chi + \sin\phi\sinh\chi\nonumber\\
\Delta_n^{AS}&=&e^{-\theta}\sin\phi\cosh\chi + \cos\phi\sinh\chi\nonumber\\
\Delta_n^{SA}&=&-e^{\theta}\sin\phi\cosh\chi +\cos\phi\sinh\chi\nonumber\\
\Delta_n^{SS}&=&e^{-\theta}\cos\phi\cosh\chi -\sin\phi\sinh\chi.
\eea
where the parameters $\chi$, $\phi$ and $\theta$ are given in terms of the couplings as:
\bea
e^\theta \cosh\chi &=& \frac{2(\Delta_n^{11}\Delta_n^{12}+ \Delta_n^{21}\Delta_n^{22})+\gamma}{2\gamma^{1/2}},\label{eq19}\\
e^{-\theta} \cosh\chi &=& \frac{-2(\Delta_n^{11}\Delta_n^{12}+\Delta_n^{21}\Delta_n^{22})+\gamma}{2\gamma^{1/2}}\label{eq20},\\
\gamma&=&(\Delta_n^{12}-\Delta_n^{21})^2 +(\Delta_n^{11}+\Delta_n^{22})^2.
\eea
We have seen how to obtain a Hamiltonian equivalent to Harper's model in an abelian sector of the theory. As presented in previous sections, this produces indeed a butterfly coming from an AMO. 


On the other hand, non-negligible elements in $\delta_{SA}$ reveal the non-abelian nature of the field as we shall explore in the next section.

\section{Emulation of  SU(2) field\label{sec5}}

The present section shows that the geometric dependence of couplings can be decomposed into a function of the distance between dimer centers and a function of each dimer orientation. The latter dependence is crucial in the construction of arrays with localized wave functions as the one presented in section \ref{sec5.3}.

\begin{figure*}[t]
\includegraphics[width=0.6\textwidth]{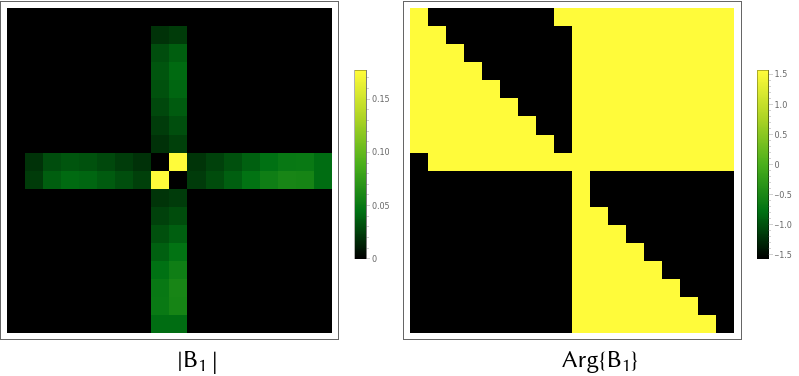}
\caption{Component $\sigma_1$ of the field $B_z$. This field is the result of the commutator $[A_x,A_y]$ in eq. (\ref{eq53}) and constitutes the purely non-abelian contribution of the field. The other components of $B_z$ are small in comparison.  }
\end{figure*}

\begin{table}[b]
\caption{\label{tablefractal} Fractal dimensions of the symmetric mode band.}
\begin{ruledtabular}
\begin{tabular}{cc}
$\omega 2 \pi/(N-1)$ & Fractal dimension\\\colrule
0                    &  0.9606          \\ 
1                    &  0.9601          \\
2                    &  0.4980          \\
3                    &  0.4884          \\
5                    &  0.4244          \\
7                    &  0.6525          \\
11                   &  0.6848          \\
13                   &  0.6995          \\
17                   &  0.6941          \\
19                   &  0.7010          \\
23                   &  0.7233          \\
29                   &  0.6897          \\
31                   &  0.7308          \\
37                   &  0.7516          \\
41                   &  0.7389          \\
\end{tabular}
\end{ruledtabular}
\end{table}

\subsection{A model with SU$(2)$\label{sec5.1}}




Following the general formalism exposed in sections \ref{sec3.1} and \ref{sec3.2}, we study the construction of a square lattice made of dimers.  We define the projectors $P_\pm=(1\pm \sigma_3)/2$ that select the state 1 or 2 inside one dimer.  The coupling between state 1 of a lattice site with state 2 belonging to its neighbor at the right would correspond to the operator $T_1 \sigma_{+}$ and its hermitian conjugate, by reciprocity. A square lattice with nearest lattice couplings, internal dimer couplings and arbitrary on-site energies is described by the following Hamiltonian 
\bea
H &=& (\Delta_{1}^{+} \sigma_{+} + \Delta_{1}^{-} \sigma_{-}+ \tilde \Delta_{1}^{+} P_{+} + \tilde \Delta_{1}^{-} P_{-}) T_1 + h.c. \nonumber \\
&+& (\Delta_{2}^{+} \sigma_{+} + \Delta_{2}^{-} \sigma_{-}+ \tilde \Delta_{2}^{+} P_{+} + \tilde \Delta_{2}^{-} P_{-}) T_2 + h.c. \nonumber \\
&+& \Delta_{0}^{+} \sigma_{+} + \Delta_{0}^{-} \sigma_{-} +\tilde \Delta_{0}^{+} P_{+} + \tilde \Delta_{0}^{-} P_{-} + V
\label{1.18}
\eea
whereas a triangular lattice (which has coordination number 6) made of dimeric sites has a Hamiltonian
\bea
H &=& ( (\Delta_{1}^{+} \sigma_{+} + \Delta_{1}^{-} \sigma_{-}+ \tilde \Delta_{1}^{+} P_{+} + \tilde \Delta_{1}^{-} P_{-}) T_1 + h.c. \nonumber \\
&+& (\Delta_{2}^{+} \sigma_{+} + \Delta_{2}^{-} \sigma_{-}+ \tilde \Delta_{2}^{+} P_{+} + \tilde \Delta_{2}^{-} P_{-}) T_2 + h.c. \nonumber \\
&+&(\Delta_{3}^{+} \sigma_{+} + \Delta_{3}^{-} \sigma_{-}+ \tilde \Delta_{3}^{+} P_{+} + \tilde \Delta_{3}^{-} P_{-})T^{\dagger}_2 T_1 + h.c. \nonumber \\
&+& \Delta_{0}^{+} \sigma_{+} + \Delta_{0}^{-} \sigma_{-} +\tilde \Delta_{0}^{+} P_{+} + \tilde \Delta_{0}^{-} P_{-} + V
\label{1.19}
\eea
where $T_3$ is actually $T^{\dagger}_2 T_1$. The six neighbors are obtained here if we recognize that the hermitian conjugate of the third row reverses the sign of the vector $\v a_1-\v a_2$. The Hamiltonian of a particle in a nearest neighbor lattice of coordination number $c$ can be written in the same fashion. Moreover, we choose to express $\sigma_{\pm}$ in terms of $\sigma_{1,2}$, $P_{\pm}$ in terms of $\sigma_3$ and $\sigma_0 = \v 1$. The resulting expression is refreshingly simple if we adopt the Einstein convention for greek indices in $\sigma_{\mu}$, \ie
\bea
H = \sum_{k=1}^{c} \Delta_{k}^{\mu} \sigma_{\mu} T_k + \Delta_0^{\mu} \sigma_{\mu} + h.c. ,
\label{1.20}
\eea 
albeit no Lorentzian metric has been required for this theory.

If the operators $\Delta_k^{\mu}$ representing local-dependent couplings are all parallel, then $\Delta_k^{\mu}=\Delta^{\mu}$ for all $j$, and the Hamiltonian (\ref{1.20}) will represent a lattice with site-dependent deformations but without a gauge field. We have the following factorization property:
\bea
H = \Delta^{\mu} \sigma_{\mu} \sum_{k=1}^{c}  T_k  + h.c. + \Delta_0^{\mu} \sigma_{\mu}
\label{1.21}
\eea
Another case of reduction is the decoupling of identical lattices, \ie $\Delta_k^{1} = \Delta_k^{2} = \Delta_k^{3} =0$ for all $k$, leading to 
\bea
H= \v 1 \otimes \left\{ \sum_k \Delta^0_k T_k + h.c. + V \right\},
\label{1.22}
\eea
\ie a theory without internal structure. It is important to note that in (\ref{1.20}), the coupling constants are complex in general; however, in specific realizations, the use of real and positive $\Delta_{k}^{\pm}$ for each site shall impose some restrictions on the values of $\Delta_{k}^{\mu}$. Evidently, the result is aperiodic in general, as described in Fig. \ref{fig:linearchain}. 



Before diagonalizing each dimer, we decompose each site-to-site coupling in center-to-center couplings plus fluctuations:
\bea
\Delta_k^\mu(d_{n,m}^{i,j}) = \Delta(L) + \delta_k^\mu(d_{n,m}^{i,j})
\label{eq27}
\eea 
The distances $d_{n}^{i,j}$ explained in Fig. \ref{pairwisecoup}, can be expressed in terms of reference lengths $L, d$  and the angles $\theta_n, \theta_{n+1}$, such that the dependence of $\delta_k^\mu$ on such parameters can be exhibited. Assuming only orientational dependence but no deformation, and the fact that $\Delta$ must be real and positive, one has $\Delta^3_k = \Delta^2_k = 0$, uniform on-site energies (\ie $V=0$ can be imposed) and the Hamiltonian becomes
\bea
H=& (\v 1 + \sigma_1) \Delta(L) \sum_k T_k + h.c. \nonumber\\
&+ \sum_k \delta^{\mu}_k \sigma_{\mu}  T_k + h.c. + \Delta(d) \sigma_1, 
\label{1.24}
\eea
where the first term represents the periodic lattice, the second term contains all orientational fluctuations (disorder could be introduced in this part) and the last term is the intra-dimer coupling contribution. Finally, we introduce a change of basis, in order to transform each dimer into an effective two-level potential well. It suffices to diagonalize $\sigma_1$, leading to the following Hamiltonian:
\bea
\tilde H &=& P_+ \left\{\sum_k(\Delta(L) + \delta_k^1 + \delta_k^0) T_k + h.c. + \Delta(d) \right\} \nonumber \\  &+& P_+ \left\{\sum_k(\delta_k^0 - \delta_k^1) T_k + h.c.  - \Delta(d) \right\}
\nonumber \\ &+& \sigma_1 \sum_k \delta_k^3 (T_k+T_k^{\dagger})
\label{1.25}
\eea
or in block form
\begin{widetext}
\bea
\tilde H = \left( \begin{array}{cc} \sum_k(\Delta(L) + \delta_k^1 + \delta_k^0) T_k + h.c. + \Delta(d)  & \sum_j \delta_k^3 (T_k+T_k^{\dagger}) \\  \sum_k \delta_k^3 (T_k+T_k^{\dagger})  & \sum_k(\delta_k^0 - \delta_k^1) T_k + h.c.  - \Delta(d) \end{array}\right)
\label{eq42}
\eea
\end{widetext}
where each block is a Hamiltonian for the global lattice. The off-diagonal blocks represent the non-abelian nature of the resulting theory, in the basis of dimer eigenfunctions. The fact that this dependence cannot be removed, indicates that the emulation of interactions with an external non-abelian field depend crucially on the orientational fluctuations $\delta_k^1, \delta_k^3$, which in turn can be geometrically designed as indicated in Fig. \ref{fig:linearchain}. 

By comparing the Hamiltonians (\ref{eq42}) and (\ref{eq30}) we can conclude that the non-abelian behavior of the field is originated by the dimer orientations of the chains.

\begin{figure}[b]
\includegraphics[width=0.4\textwidth]{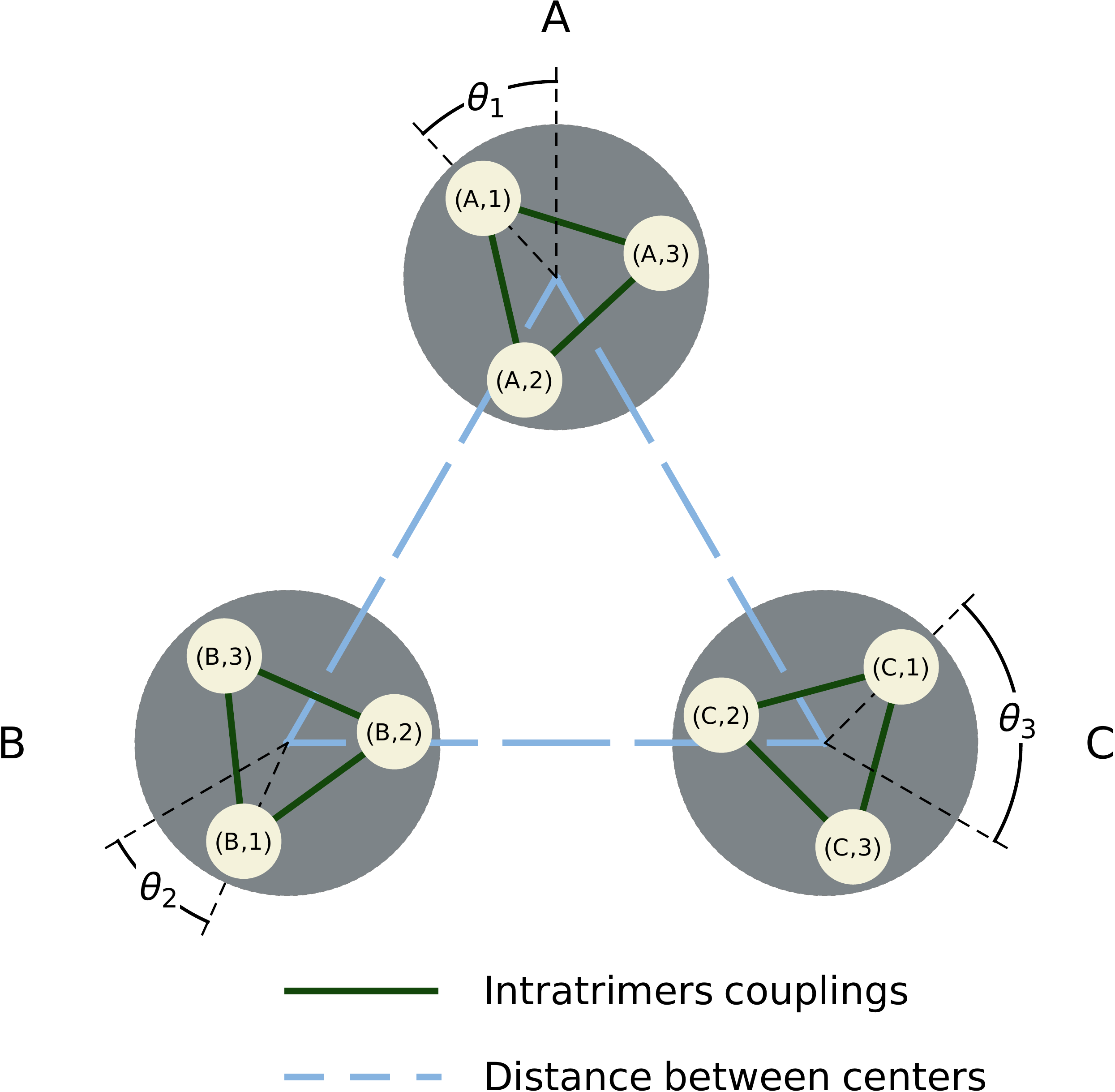}
\caption{\label{fig11} Unit cell for the chain of trimers. Each internal triangle can rotate independently with angles $\theta_1$, $\theta_2$ and $\theta_3$. The dashed blue lines represent the distances between centers. }
\end{figure}

\subsection{Effective negative couplings and level inversion revisited\label{sec5.2}}

When the fluctuations are small, we may try to find an approximate representation that eliminates the off-diagonal blocks. Here, we must recover an abelian gauge field, but by the same token, this field must be described by the fluctuations in the orientations. If we simply eliminate $\delta_j^3$, the second block has a peculiarity already noted in \cite{Sadurni_2019,yael_2020}: level inversion can be achieved whenever $\delta_j^0 < \delta_j^1$, and even a degeneracy point occurs when $\delta_j^0 = \delta_j^1$, \ie a hidden symmetry that does not pertain to the underlying lattice symmetries. The level inversion corresponds to the realization of negative couplings using exclusively positive on-site waves. Another way to look at this, according to \cite{Sadurni_2019}, is to recognize that the basis of a dimer contains both symmetric and antisymmetric functions, which means that couplings between dimeric excited states can change sign if the antisymmetric waves enter with opposite signs in the overlap integrals.

In the case where the fluctuations are indeed small, but their effect is not completely neglected to second order, a perturbative approach can help to reabsorb them in the diagonal part by a unitary transformation.

\subsection{Wave confinement in SU(2). Bound states\label{sec5.3}}

In section \ref{sec2} we showed examples of lattices that exhibit the confinement phenomenon of its wave function in the center of the array. In this section we present the construction of the corresponding square lattice.

Our method of construction is based on the intensities of the couplings between two adjacent dimers in the direction of the primitive vectors. The strongest couplings must be located at the center of the array while the orientation of the dimers out of this region must be disposed in a configuration that favors small coupling values.  We achieve these conditions with four central  dimers illustrated in Fig. \ref{fig4}. In general, the position of each site is given by the following expressions:
\begin{widetext}
\bea
\mathbf{a_1^{(\v n)}}&=&\left((n_x-1)l-d\sin\left[\frac{\pi}{4}+\frac{\pi}{2(n-2)} \right](n_y -n_x) ,-(n_y-1)l-d\cos\left[\frac{\pi}{4}+\frac{\pi}{2(n-2)} \right](n_y -n_x) \right)\\
\mathbf{a_2^{(\v n)}}&=&\left((n_x-1)l-d\sin\left[\frac{5\pi}{4}+\frac{\pi}{2(n-2)} \right](n_y -n_x) ,-(n_y-1)l-d\cos\left[\frac{5\pi}{4}+\frac{\pi}{2(n-2)} \right](n_y -n_x) \right),
\eea
\end{widetext}
where $\v a_1^{(\v n)}$ and $\v a_2^{(\v n)}$  stand for the positions 1 and 2 of the dimer at $\v n =(n_x,n_y)$. Now we vary the lattice parameter and find a localized stated at the top of the energy band. This localized state moves downwards, as the lattice constant increases with fixed dimeric distance. In fact, it merges with the propagation band in the limit of negligible intradimeric distance --small compared to the lattice constant-- and reveals the non abelian nature of the effect. An analogy with solitons can be made here, as the localized state dwells in a region produced by a Hedgehog-type external field. This configuration of dimers has some characteristics in common with solutions of the classical non-linear field equations possessing a finite energy and localized non-dispersive energy density \ie this wave travels with its shape unchanged\cite{rajaraman,scott1973}. Moreover, such solution falls to zero at infinity. In the non-linear theory we pretend to emulate, the solutions perturb our discrete Schr\"odinger equation. 

The portrait of the wave function for the higher energy state was shown in section \ref{fig4} in Fig. \ref{sqrlattice}. We observe that the electronic state is confined at the center of the array, but we are also interested in the shape of the emulated non-abelian field.  In the direction perpendicular to the lattice plane, $B_z$ is obtained as follows:
\be\label{eq53}
B_z=\partial_y A_x -\partial_x A_y-i[A_x,A_y],
\ee
where the commutator plays here an important role. In order to calculate the components of $A$, first we must express each $2\times 2$ coupling block as an exponential:
\be
\Delta(\v N)=\exp(\bfPhi^{(0)}\v 1+\bfPhi\cdot \bfsigma).
\ee
The component $A_i$ is the difference between $\bfPhi(\v N)$ and $\bfPhi(\v N-\hat{\v e}_i)$. The projections on the Pauli basis $\sigma_0, \sigma_1, \sigma_2$ and $\sigma_3$ of $\v A$ are readily found by computing Tr$(A \sigma_i)$ and are illustrated in Fig \ref{afield1}. The real part of $A^{(0)}$, $A^{(2)}$ and the imaginary part of $A^{(1)}$, $A^{(3)}$ are zero.

\begin{figure*}
\begin{tabular}{cc}
\includegraphics[width=0.4\textwidth]{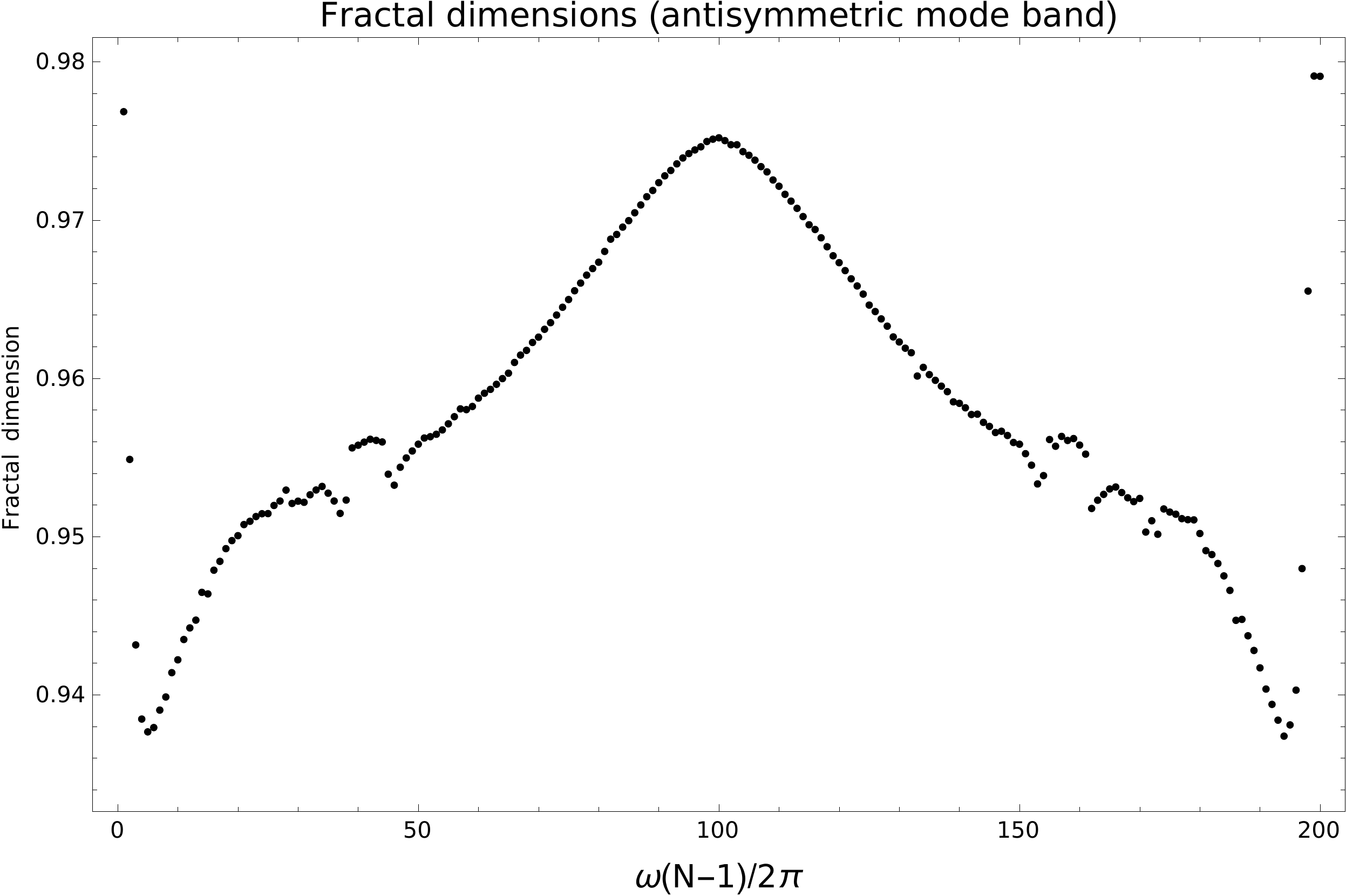}&\includegraphics[width=0.4\textwidth]{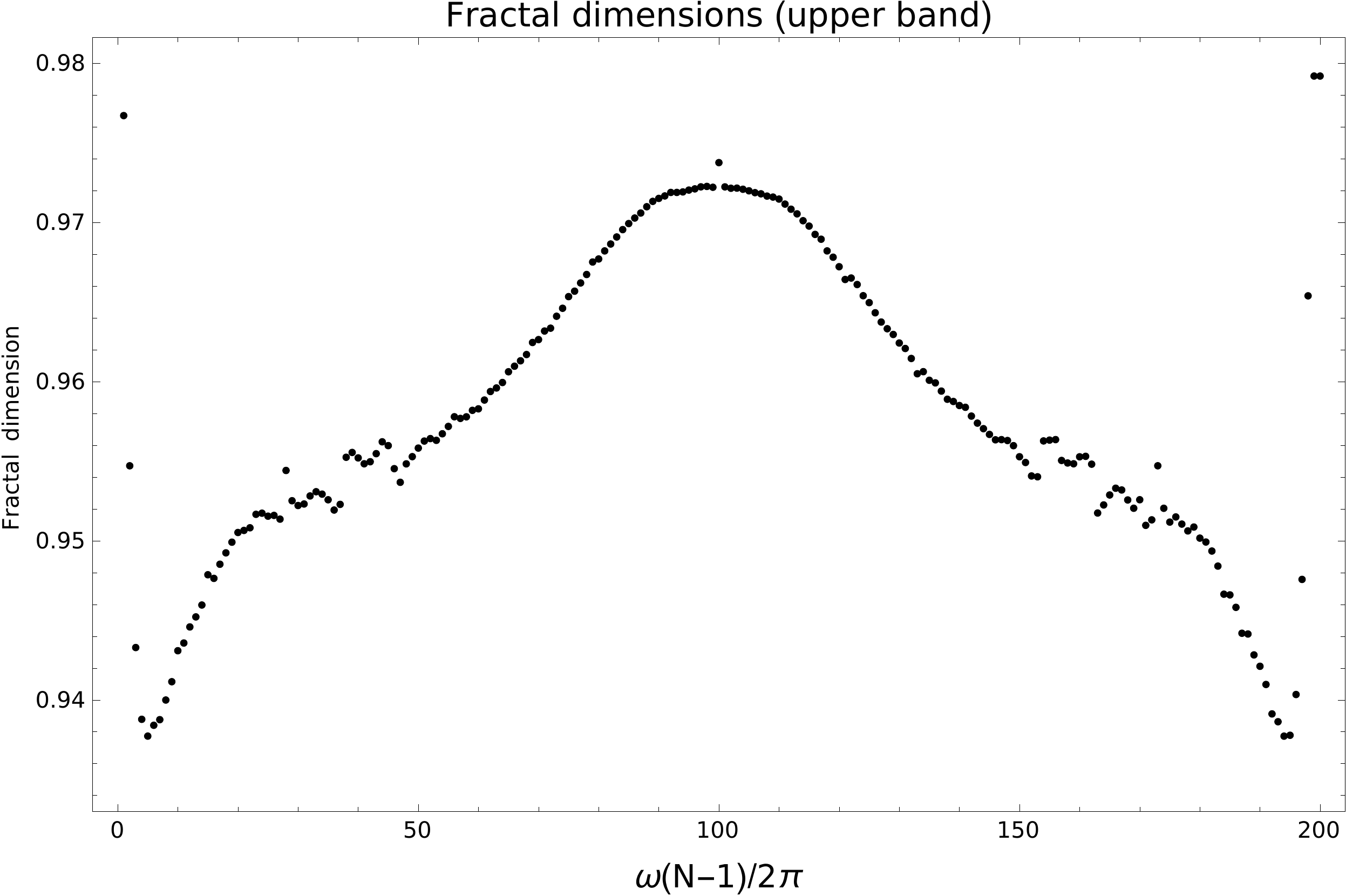}\\
\\
{\large (a)}&{\large (b)}\\
\end{tabular}
\caption{\label{fracdim} Panels (a) and (b) present the fractal dimensions of the antisymmetric mode band and upper band respectively. Both spectra show similarities in the fractal dimension values. 
}
\end{figure*}

\section{Band mixing in SU(3) models with color\label{sec6}}

The emulation of SU(3) fields can be performed in a similar way as presented for the previous SU(2) case. This time we construct a liner chain made of trimers, where the unit cell is shown in Fig. \ref{fig11}. The energy levels of each internal trimer represent three internal degrees of freedom in the cell. As the  triangles rotate independently with angles $\theta_1, \theta _2$ and $\theta_3$, we can break or restore the local polygonal symmetry of each cell, and thus manipulate the emergent band structure. When these angles vanish, the system has $C_3\times C_3$ symmetry.  The Hamiltonian of the array is, in matrix form: 
\bea
H&=&
\left(
\begin{array}{ccc}
H_0&\Delta_{AB}&\Delta_{AC}\\
\Delta^\dagger_{AB}&H_0&\Delta_{BC}\\
\Delta^\dagger_{AC}&\Delta^\dagger_{BC}&H_0
\end{array}
\right),\\
H_0&=&
\left(
\begin{array}{ccc}
E&\Delta&\Delta\\
\Delta&E&\Delta\\
\Delta&\Delta&E
\end{array}
\right),\\
\Delta_{mn}&=&
\left(
\begin{array}{ccc}
\Delta_{m1,n1}&\Delta_{m1,n2}&\Delta_{m1,n3}\\
\Delta_{m2.n1}&\Delta_{m2,n2}&\Delta_{m2,n3}\\
\Delta_{m3,n1}&\Delta_{m3,n2}&\Delta_{m3,n3}
\end{array}
\right),
\eea
where $m,n=A,B,C$ and $\Delta_{ml,nk}$ is the coupling between the $l$-th site of the trimer $m$ with the $k$-th site of trimer $n$. A transformed Hamiltonian is obtained if we use the $3\times 3$ matrix that diagonalizes $H_0$, i.e. $\tilde{H}=U^\dagger H_0 U$ where $U=\mathbf{1}\otimes U_{3\times 3}$. The transformed matrix elements are calculated analytically below: 
\bea
\left(U_{3\times 3}\right)_{p,q}&=&\frac{1}{\sqrt{3}}\exp\left[\frac{i2\pi}{3}(2+p)q \right],\\
\left(\tilde{\Delta}_{mn}\right)_{pq}&=&\frac{e^{4\pi i (q-p)/3}}{3}\sum_{k,k'=1}^3 e^{2\pi i(k'q-kp)/3}\Delta_{mk,nk'}.\nonumber\\
\eea
Now, the Hamiltonian is expressed in the eigenbasis of $C_3$: Reorganizing matrix elements leads to a Hamiltonian dived into three blocks: 
\begin{widetext}
\bea\label{eq:hmatrix}
\small
H=\left(
\begin{array}{ccccccccc}
S_1&(\Delta'_{AB})_{33}&(\Delta'_{AC})_{33}&0&(\Delta'_{AB})_{31}&(\Delta'_{AC})_{31}&0&(\Delta'_{AB})_{32}&(\Delta'_{AC})_{32}\\
(\Delta'_{AB})^*_{33}&S_2&(\Delta'_{BC})_{33}&(\Delta'_{AC})^*_{13}&0&(\Delta'_{BC})_{31}&(\Delta'_{AC})^*_{23}&0&(\Delta'_{BC})_{32}\\
(\Delta'_{AC})^*_{33}&(\Delta'_{BC})^*_{33}&S_3&(\Delta'_{AC})^*_{13}&(\Delta'_{BC})^*_{13}&0&(\Delta'_{AC})^*_{23}&(\Delta'_{BC})^*_{23}&0\\ 
0&(\Delta'_{AB})_{13}&(\Delta'_{AC})_{13}&D_{11}&(\Delta'_{AB})_{11}&(\Delta'_{AC})_{11}&0&(\Delta'_{AB})_{12}&(\Delta'_{AC})_{12}\\
(\Delta'_{AB})	^*_{31}&0&(\Delta'_{BC})_{13}&(\Delta'_{AB})^*_{11}&D_{21}&(\Delta'_{BC})^*_{11}&(\Delta'_{AB})^*_{21}&0&(\Delta'_{BC})_{12}\\
(\Delta'_{AC})^*_{31}&(\Delta'_{BC})^*_{31}&0&(\Delta'_{AC})^*_{11}&(\Delta'_{BC})^*_{11}&D_{31}&(\Delta'_{AC})^*_{21}&(\Delta'_{BC})^*_{21}&0\\ 
0&(\Delta'_{AB})_{23}&(\Delta'_{AC})_{23}&0&(\Delta'_{AB})_{21}&(\Delta'_{AC})_{21}&D_{12}&(\Delta'_{AB})_{22}&(\Delta'_{AC})_{22}\\
(\Delta'_{AB})^*_{32}&0&(\Delta'_{BC})_{23}&(\Delta'_{AB})^*_{12}&0&(\Delta'_{BC})_{21}&(\Delta'_{AB})^*_{22}&D_{22}&(\Delta'_{BC})_{22}\\
(\Delta'_{AC})^*_{32}&(\Delta'_{BC})^*_{32}&0&(\Delta'_{AC})^*_{12}&(\Delta'_{BC})^*_{12}&0&(\Delta'_{AC})^*_{22}&(\Delta'_{BC})^*_{22}&D_{32}
\end{array}
\right).
\eea
\end{widetext}
The symmetry breaking of a global $C_3$ of each trimer is introduced by a rotation of only one of the internal dimers; as a consequence, triple degeneracy points emerge as announced previously. Note though that doublets are never split.  The internal $C_3$ breaking is achieved by deformation  of the internal trimers, and in this case lifted doublets appear. Following the same path as in the case of SU(2), we construct a linear chain made of rotated unit cells. This system produces the spectrum of the Fig. \ref{figtrimers} where we find overlapped bands.  The superior band contains nine copies of the Hofstadter butterfly. The model shows emergent flat bands, of potential interest in recent studies on superconductors.

Finally, the model can be generalized also to a 2D lattice, in a similar way to the SU(2) case, with the aim of emulating confinement due to SU(3) color fields.

\section{Conclusion\label{sec7}}

In this work, the emulation of non abelian fields was studied in periodic, quasi-periodic and defective lattices with internal degrees of freedom given by the energy levels of a polymer. The full lattice is homogeneous, i.e. all sites were identical. For a given set of geometric parameters in a linear chain of rotating dimers, the projections of the energy levels in the uncoupled symmetric and antisymmetric states reproduced well Harper's Hamiltonian spectra with $\Lambda=1.7$ and $\Lambda=1$ for the two emergent bands. It was also possible to show that the spectrum was independent of Bloch' quasimomentum, as in the standard theory of ergodic operators. This gave us a clear indication of an emergent magnetic field produced by mere deformations. 

In the general description of coupled bands, the orientations of the couplings between neighboring sites were responsible for non zero elements in the diagonal block of the Hamiltonian; this showed that our construction was a genuine realization of a non-abelian theory in a regime of strong couplings between bands. The phenomenon of confinement due to SU(2), was emulated in a square array with dimers where the hopping amplitudes were tuned in order to produce an effective (non-diagonal) defect; the wave function as shown to be located at the center of the crystal and its energy was found on the superior part of the propagation band. This localized state merged with the latter, as the lattice constant increased, which resulted in a fully periodic array in the limit of negligible dimer size. It is important to note that the dimeric structure could be shown to be indispensable to determine whether a non-abelian group SU(2) entered or not in the description. The bound state located at the lattice defect would simply disappear in the absence of a non-abelian theory. 

A few words are in order in connection with technological applications. The magnetic field emulated with deformations allows to study similar effects in mesoscopic systems and wave transport. Its applications to wave propagation in elastic tight-binding emulations \cite{ramirez2020} and photonic crystals offers itself as a type of metallic-to-insulator transition in systems where external fields that manipulate waves are not available.
We have also shown that the introduction of extra orbitals in solid state physics leads to deformations of Cantor sets. As an outlook, we plan to analyze the implications of such spectral deformations on the Quantum Hall effect \cite{miransky_quantum_2015,thouless_quantized_1982}, whose mysteries are yet to be fully explained. 

\begin{acknowledgments}
Beca 291197 CONACYT and project 100518931-VIEP BUAP-CA-289.
\end{acknowledgments}

\appendix

\section{Unitary map between translation operators in 1D; beyond Aubry's duality. }
\label{appendixa}
In this section we show explicitly the unitary map of the Harper Hamiltonian into a tight-binding operator without potential, but with variable couplings. For ease in the calculations, we shall work in a representation where the transformation an site number operators act on smooth functions, \ie $T\rightarrow e^{iap}$, $p=-i\hbar\partial_x $, $N\rightarrow x/a$. If we act on a complete basis of Wannier functions $\phi_b(x-a)$ this represents no loss of generality. However, we must make sure that our unitary map takes us from periodic functions (period $a$, index band $b$) to periodic functions. Let us take the following unitary transformation on the Hamiltonian
\be
\tilde{H}=UHU^\dagger,\quad UU^\dagger=\mathbf{1}.
\ee
The unitary operator $U$ can be expressed as:
\be
U=V(N)\exp\{i(\alpha p^2 +\beta p) \},
\ee
such that $V(N)V^\dagger(N)=\mathbf{1}, \alpha , \beta \in \rcal$. We shall see that $U$ does not change the nature of the lattice if $\alpha$ and $\beta$ are properly chosen. We have 
\be
T=e^{iap}, \quad E\equiv e^{i\omega N}=e^{i\omega x/a} 
\ee
and defining $v\equiv \omega /a$, $E$ can be written as $E\equiv e^{ivx}$, where $x=i\hbar \partial_p$. With these definitions, the Harper's Hamiltonian (Almost Mathieu Operator) has the form:
\be
H=T+T^\dagger + \Lambda \left(E+E^\dagger \right),
\ee
for $\nu = 0$, which is  isospectral to any $\nu\neq 0$. Direct substitution leads to 
\bea
UTU^\dagger&=&V(N)e^{i(\alpha p^2 +\beta p)}Te^{-i(\alpha p^2 +\beta p)}V^\dagger (N)\nonumber\\
&=&V(N)TV^\dagger (N) = V(x/a)TV^\dagger (x/a)\nonumber\\
&=&V(N)V^\dagger (N+1)T.
\eea 
In addition 
\be
UEU^\dagger = V(N)e^{i(\alpha p^2 +\beta p)}Ee^{-i(\alpha p^2 +\beta p)}V^\dagger (N)
\ee
and 
\be
Ef(p)=f(p-v)E,
\ee
so that
\bea
UEU^\dagger &=& V(N)e^{i(\alpha p^2 +\beta p)}e^{-i[\alpha (p-v)^2 +\beta (p-v)]}E V^\dagger (N)\nonumber\\
&=&V(N)e^{i(2\alpha v p +\beta v -\alpha v^2)}E V^\dagger (N)\nonumber\\
&=&e^{i(\beta v -\alpha v^2)}V(x/a)e^{i2\alpha v p}EV^\dagger (x/a)\nonumber\\
&=&e^{i(\beta v -\alpha v^2)}V(x/a)e^{i2\alpha vp}EV^\dagger (x/a)\nonumber\\
&=&e^{i\beta v -\alpha v^2}V(x/a)V^\dagger \left(\frac{x+2\alpha v}{a}i\right)e^{iqap}E.
\eea
This is a closed result within the space of lattice translations if $2\alpha v=qa,\ q\in \zcal$. The transformed $E$ is:
\bea
UEU^\dagger &=& e^{i\beta v - \alpha v^2} V(N)V^\dagger (N+q) T^q E\nonumber\\
&=&e^{i\beta v -\alpha v^2}V(N)V^\dagger (N+q) ET^qe^{ivaq}\nonumber\\
&=&e^{i(\beta v -\alpha v^2 +vaq)}V(N)V(N+q)ET^q\nonumber\\
&=&e^{i(\beta v -qav/2 +vaq)}V(N)V^\dagger (N+q)ET^q\nonumber\\
&=&e^{i\chi_q}V(N)V(N+q)ET^q.
\eea
Therefore, the new Hamiltonian becomes:
\be
\tilde{H} = V(N)V^\dagger (N+1)T+\Lambda e^{i\chi_q}V(N)V^\dagger(N+q)ET^q + h.c.
\ee
which is a multiple neighbor tight-binding operator. Since the parameters in $U$ can chosen at will, we have that for $q=1$ or $q=-1$, $H$ reduces to a nearest-neighbor Hamiltonian. Finally, we recognize that every nearest-neighbor tight-binding model in 1D can be cast in the form of a Hamiltonian with real and positive couplings by means of gauge transformation. In general, we have 
\be
\Delta=V(N)V^\dagger (N+1)\left(1+\Lambda e^{\pm i\chi_q}e^{\pm i\omega N}\right).
\ee
We recognize the unitary nature of $V(N)$ so $VV^\dagger = \mathbf{1}$ implies $V=\exp(i\Phi (N))$ and $V(N)V^\dagger(N+1)=\exp\{i(\Phi(N)-\Phi(N+1)\}\equiv \exp(-i\delta\Phi(N)) \}$ which is only a phase factor, and then
\bea
\Delta=e^{i\Omega(N)}|\Delta(N)|, \ \tilde{U}=e^{i\sum_{j=-\infty}^n \Omega(N+j)}\\
T\tilde{U}^\dagger=e^{-i\sum_{j=\infty}^{n+1}\Omega(N+j)i}T\\
\tilde{U}T\tilde{U}^\dagger= e^{-i\Omega(N+n+1)}T.
\eea
If we choose $n=-1$ we obtain the following expression for $|\Delta|$
\be
|\Delta|=\sqrt{1+\Lambda^2+2\Lambda\cos(\omega N+\chi_\pm)},
\ee
which is the sought expression. Geometric deformations are therefore equivalent to on-site quasiperiodic energies.

In the case of a very faint intensity of the potential, we can analyze the limit $\Lambda\ll 1$ and provide an expression for the coupling:
\be\label{eq:a16}
|\Delta|\simeq 1+\Lambda\cos(\omega N + \chi_\pm),\quad \Lambda\ll 1,
\ee 
which is in full agreement with the first corrections to the effective coupling in the problem of rotating dimers.

\section{Box-counting fractal dimension calculation}
The fractal dimensions of the symmetric and antisymmetric spectra of the chain made of dimers were calculated using the method of box counting:
\be
N(s)=\left(\frac{1}{s} \right)^D,
\ee
where $N$ is the number of boxes and $s$ is the size of one box. Thus the fractal dimension can be calculated as follows:
\be
D=\frac{\log N(s)}{\log(1/s)}.
\ee
The fractal dimensions for specific values of $\omega(N-1)/2\pi$ of the spectrum showed Fig. \ref{fig1} (b) are presented in Fig. \ref{fracdim}. The range of values of the fractal dimensions for the symmetric mode band are reported in Table \ref{tablefractal}.

\end{document}